\documentclass[aps,twocolumn,prx,amsmath,amssymb,showpacs,floatfix,altaffilletter]{revtex4-2}

\usepackage{graphicx}
\usepackage{changes}
\usepackage{times}
\usepackage[varg]{txfonts}
\usepackage{bm}
\usepackage{amsmath}
\usepackage{sansmath}
\usepackage{xspace}
\usepackage{xr}
\usepackage{natbib} 
\usepackage{xcolor}
\usepackage[caption=false]{subfig}
\usepackage{appendix}
\usepackage{longtable}
\usepackage{tabularx}
\usepackage{setspace}
\usepackage{booktabs}
\usepackage{braket}
\usepackage{csquotes}
\usepackage{wrapfig}
 
\usepackage{todonotes} 
\usepackage{array}

\usepackage[colorlinks,bookmarks=false,citecolor=blue,linkcolor=red,urlcolor=blue]{hyperref}
\usepackage{cleveref}

\newcommand{\beq}{\begin{equation}}
\newcommand{\eeq}{\end{equation}}

\begin{document}

\title{The landscape of symmetry enhancement in tight-binding models}

\author{A. K. Dagnino$^{1,2}$, A. Corticelli$^{2}$, M. Gohlke$^{3}$, A. Mook $^{4}$, R. Moessner$^{2}$, P. A. McClarty$^{2,5}$}
\affiliation{$^{1}$ Department of Physics, University of Z\"{u}rich, Winterthurerstrasse 190, CH-8057 Z\"{u}rich, Switzerland}
\affiliation{$^{2}$ Max Planck Institute for the Physics of Complex Systems, N\"{o}thnitzer Str. 38, 01187 Dresden}
\affiliation{$^{3}$ Theory of Quantum Matter Unit, Okinawa Institute of Science and Technology Graduate University, Onna-son, Okinawa 904-0495, Japan}
\affiliation{$^{4}$ Institute of Physics, Johannes Gutenberg University Mainz, 55128 Mainz, Germany}
\affiliation{$^{5}$ Laboratoire L\'{e}on Brillouin, CEA, CNRS,  Universit\'{e} Paris-Saclay, CEA-Saclay, 91191 Gif-sur-Yvette, France}

\begin{abstract}
Band structures are ubiquitous in condensed matter physics and their symmetries constrain possible degeneracies, topology and response functions across a broad range of different systems. Here we address the question: given a parent crystal, what is the symmetry of hopping models on that lattice at a given shell number? We find that the parent structure does not, in general, determine the symmetry of the tight-binding model. Instead, the symmetry is dependent on the hopping range. The key to symmetry breakdown on the lattice is the existence of different {\it bond equivalence classes} whose number is related to group-subgroup indices for a broad classes of cases. We find all bond equivalence classes for $s$-wave hopping out to 20th neighbor across the different space groups and Wyckoff positions and the symmetries of the associated tight-binding models. 
These observations naturally lead to the definition of a {\it bond complex} $-$ the possible classes of networks of bonds to which symmetries may be enhanced from a given parent structure. 
\end{abstract}

\maketitle

\tableofcontents

\section{Introduction}
\label{sec:introduction}

A beginning student of condensed matter does not have to wait long to be introduced to the problem of determining the spectrum of particles hopping on a lattice. This problem is well-motivated as a tight-binding approximation to the electronic band structures of crystalline solids. But students will also appreciate that similar considerations apply to band structures of phonons and magnons. Indeed, this list can be extended -- photonic crystals,  particles propagating in optical lattices, $\ldots$ -- essentially {\it ad nauseam}: few concepts in physics are similarly ubiquitous.

In recent developments, this concept has taken a particularly prominent role: the advent of topological band structures has led to studies of the properties of hopping problems far beyond determining their spectra and dispersion relation. Most saliently, topological properties of mappings implied by the band structure have taken center stage. 

This has led to an avalanche of interest, culminating in the marriage of notions of symmetry and topology \cite{watanabe2015,po2016,watanabe2016,bradlyn2017,cano2018,watanabe2018,song2018a,song2018b} in the classification of band structures which ties also to materials through so-called topological quantum chemistry \cite{bradlyn2017,vergniory2019,topmat}. The goal of this program was to provide a complete and exhaustive classification of all possible band structures realised on regular lattices. Building on the classification of all space groups, this program has been accomplished also for magnetic symmetry groups \cite{elcoro2020magnetic,xu2020}. 

Physicists are well acquainted with the idea of the symmetry of a problem being determined by the energy scale at which experiments are done. At the level of crystallographic symmetries, the magnetic space groups implicitly assume strong spin-orbit coupling. But in the limit where spin-orbit is switched off, these symmetries groups are enlarged to the set of spin-space groups by allowing for various degrees of coupling of symmetry transformations in real and spin space \cite{BrinkmanElliott1966,BrinkmanElliott1966b,Brinkman1967,spinGroupsLO,spinPointLitvin,corticelli2021spinspace,Liu,schiff2023,xiao2023spin,ren2023enumeration,jiang2023enumeration}. One currently highly topical instance of the application of such ideas is in the study of altermagnetism, billed as a class of materials that blends features commonly associated with both ferro- or antiferromagnetism \cite{hayami2019,Smejkal2020,Smejkal,mcc2024}. Experimental candidates are far from rare and cases of particular current interest include MnTe \cite{mntegonzalez2023,MnTeKrempasky2024,MnTe2Lee024}, CrSb \cite{reimers2023} and Mn$_5$Si$_3$ \cite{reichlova2024observation}.  

Although spin-orbit coupling is present generically in materials, the enhancement of magnetic symmetries to spin-space symmetries is, in practice, very common; indeed, so common that it is barely explicitly remarked upon. The most-studied magnetic models, such as Heisenberg or XY models, tend to have higher symmetries than those implied by magnetic symmetry groups of the lattices they reside on at least when the magnetism is collinear. The fact that these enhanced symmetries are only approximate is outweighed in practice by usefulness especially when the symmetry-lowering terms are, in experimental terms, barely observable.

The starting point of this work is the observation that a similar symmetry enhancement is obtained for tight-binding models in practical use even before any notion of magnetism is invoked. In numerous commonly used settings, numerical or experimental data is fit to a tight-binding model including only a subset of all possible hopping terms: {\it any} hopping problem including finite-range hoppings only {\it a priori} is not guaranteed to have the same symmetry as the set of lattice points connected by the hops nor even the set of symmetries employed to generate the hopping model. This, then, is another set of scenarios where the effective description of a condensed matter system may have higher symmetry than the microscopic description.

This includes not only all the nearest-neighbor models taught to the above-mentioned physics students. Many band structures of great current interest arise from truncated hopping problems, with flat band systems on so-called frustrated lattices being prominent examples -- and the field of kagome metals \cite{ye2018massive,park2021electronic,tan2021charge,kang2020dirac,neupert2022charge,yin2022topological,jiang2023kagome} being a particularly topical case. 

But beyond the targeted search for new exotic physics, truncated hopping models also appear in any number of bread-and-butter applications: tight-binding fits to spaghetti plots produced by density functional theory codes, or Hamiltonians reverse-engineered from fitting phonon or magnon dispersion relations to experimental datasets obtained by, say, neutron scattering techniques are truncated if only to limit the set of fitting parameters to a reasonable size.

Concretely, in this paper, we consider $s$-wave tight-binding models of the form 
\beq
H =  - \sum_{\langle i,j \rangle_n} t_{i,j} c_i^\dagger c_j + {\rm h.c.}
\eeq
corresponding to particles hopping between sites connected at shell number $n$. At a given shell number, hopping takes place between sites separated by some fixed scalar distance. So, for example, $n=1$ corresponds to hopping between nearest neighbor sites, $n=2$ between second neighbor sites and so on. The ordering of shell numbers is determined by a convention on the primitive cell edge lengths and angles that we establish below with the requirement that there is no fine-tuning of these parameters. 

We address the following problem:
\begin{quotation}
 Consider a crystal specified by a choice of symmetry data $-$ the space group and Wyckoff position (both defined below) $-$ and $s$-wave hopping on the $n$th shell of this crystal. What is the symmetry group that leaves the tight-binding Hamiltonian invariant?
 \end{quotation}
This is a rich problem: there are $230$ space groups in three dimensions each with several Wyckoff positions leading to $1732$ possible choices. Of these we consider hopping between $n$th nearest neighbors. Within this set of possible hopping models there is, of course, considerable variety. 

Our main finding is that the symmetry is frequently enhanced above that of the parent space group and we provide a complete enumeration of the tight-binding model symmetries up to $n=20$. The next section gives simple examples showing how such symmetry enhancement can take place. The existence of symmetry enhancement depending on hopping range was noted in Ref.~\cite{gohlke2023} in the context of spin wave theory where the feature was named a {\it shell anomaly}. This paper provides a classification and analysis of shell anomalies.   
Key to the enhancement of symmetry is the notion of bond equivalence classes (or bond colorings) that we informally introduce in the next section. The number of bond equivalence classes is intimately tied to the symmetry of the tight-binding model. In Section~\ref{sec:coloring} we give a precise meaning to bond equivalence classes and provide an index theorem: that the number of colors can be computed in a large class of cases from a group-subgroup index. In the remaining cases we explain the  mechanisms underlying the absence of an index theorem. 

Our exploration of all $s$-wave tight-binding models with a cutoff in the number of shells encompasses a large landscape of possible models. We survey this landscape by supplying an overview of the variety of symmetries appearing for a fixed parent group, the variation of symmetries as a function of the hopping range including how translation symmetries can be broken. We give some attention to instances where the symmetry appears never to break down to the parent symmetry regardless of the hopping range and we prove that it is indeed possible for symmetry to be enhanced for all shells. Therefore, for systems governed by $s$-wave hopping it is possible for all response functions determined by such hopping $-$ such as those directly determined by the band structure $-$ to be governed by the enhanced symmetry and not the symmetry of the crystal.  We also discuss qualitative features of the hopping that are not captured by symmetry alone including the appearance of sub-dimensional hopping. Finally we tie these general considerations to features of band structures. Our findings have direct applications across condensed matter physics wherever $s$-wave tight-binding models play a role. Moreover, they are also of interest (and use) beyond this setting. For example, the physics of the shell anomaly can also manifest itself in spontaneous symmetry breaking where metals spontaneously develop bond order, such as in kagome metals \cite{tan2021charge,park2021electronic,neupert2022charge,tazai2022charge,wagner2023phenomenology}. The concepts we develop here we of course hope are of interest (and use) beyond this setting.

\section{Overview of the paper}
\label{sec:overview}

\subsection{Pedagogical examples}

To orient the reader, we begin with two straightforward examples intended to provide some intuition for how symmetry can be enhanced above that of the parent group. These examples serve to illustrate the existence of bond equivalence classes and the presence of enhanced symmetry for tight-binding models on particular shells. In the remainder of the paper we explore these ideas in some generality by establishing the crystalline symmetries of bonds for all space groups and Wyckoff positions and for shells up to some cutoff. Many of the results are recorded for fixed shell number cutoff $n=20$ in some crystal convention detailed below. However, for cases where the symmetry does not break down to the parent symmetries in hopping models up to this $n$ we have explored the issue of convergence in more detail.

\subsubsection{Example in 2D}
\label{sssection:2D}

\begin{figure}[h!]
    \centering
    \includegraphics[width=0.8\columnwidth]{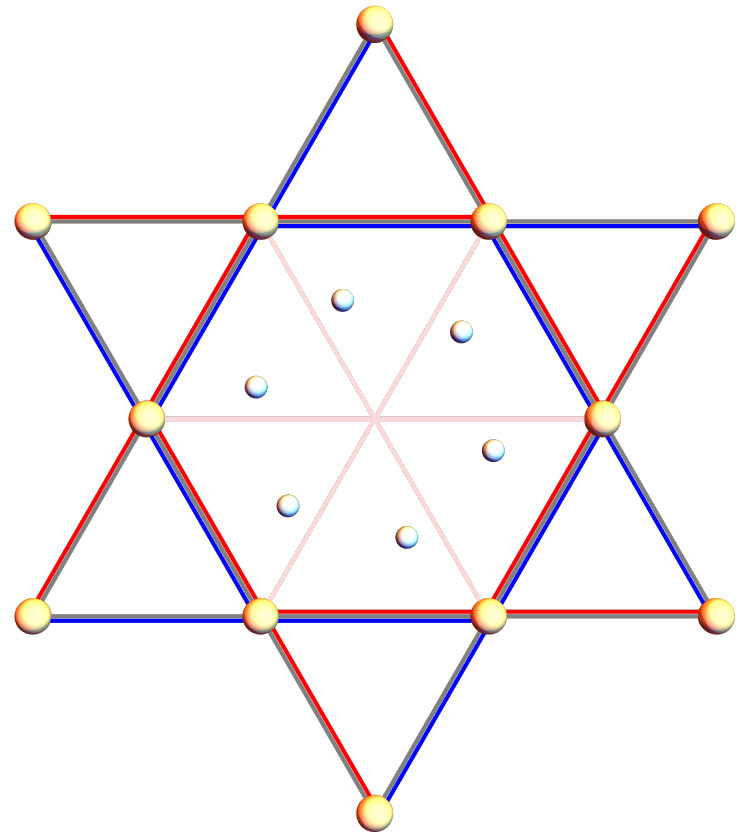}
    \caption{Illustration of the kagome lattice, whose sites are shown as yellow spheres. The white spheres reduce the point-group symmetry of the lattice from that of the undecorated kagome lattice to just $C_6$. At the third shell, there are three inequivalent classes of bonds, shown in light pink, red and blue.}
    \label{fig:kagome}
\end{figure}

Our first and simplest example is based around a decorated kagome lattice shown in Fig.~\ref{fig:kagome}. The underlying lattice is defined to have only a six-fold rotation symmetry about an axis passing through the centres of hexagons and perpendicular to the kagome plane. This is accomplished by decorating the kagome lattice (yellow vertices) with sites marked in gray, so as to reduce the overall symmetry of the crystal. Now we consider the nearest neighbor bonds connecting the orange vertices. Using the six-fold rotation symmetry together with lattice translations it is evident that all nearest neighbor bonds are equivalent to one another. So, a tight-binding model formulated with only nearest neighbor hoppings cannot know about the decoration of the kagome lattice and therefore has all the symmetries of the undecorated kagome including inversion symmetry, in-plane two-fold rotation axes and mirrors that are missing from the full crystal's symmetries. This provides a simple illustration of how symmetries can be enhanced above those of the parent crystal. 

Now we examine the third neighbor shell on the kagome lattice. These include bonds that connect vertices on opposite sides of each hexagonal plaquette on the lattice. The six-fold rotation symmetry of the parent lattice and translations ensure that all such bonds are equivalent similar to the case of nearest neighbor bonds. In Fig.~\ref{fig:kagome} these are indicated as light pink lines. But there is another class of third neighbor bonds i.e. extending over the same distance as those already considered. These bonds, shown in blue and red, stretch along nearest neighbor directions with twice the distance of the nearest neighbor bonds. It is straightforward to see that these bonds are {\it not} all equivalent. Fig.~\ref{fig:kagome} shows that the bonds colored in red map into one another under the six-fold rotations while there is no operation in the parent symmetry group that maps the red bonds into those in blue. We say that the third neighbor shell on the lattice has three bond equivalence classes. Two of these classes (or colors) are evident from the kagome itself while one class splits into two classes once the parent symmetry, translation and six-fold rotation only, is considered. 

We may now address the symmetry of the third neighbor shell. This shell has an inversion symmetry in addition to the elements of the symmetry group of the parent lattice. But it does not have the full symmetry of the kagome lattice. 

In summary, we have started with a parent crystal of low symmetry and shown that the nearest neighbor bonds have very much enhanced symmetry with a single equivalence class and that the third shell has intermediate symmetry coming from the appearance of three bond equivalence classes (of which two appear as a direct consequence of the parent lattice symmetry constraints).

\subsubsection{Example in 3D}

\begin{figure}[h!]
    \centering
    \includegraphics[width=\columnwidth]{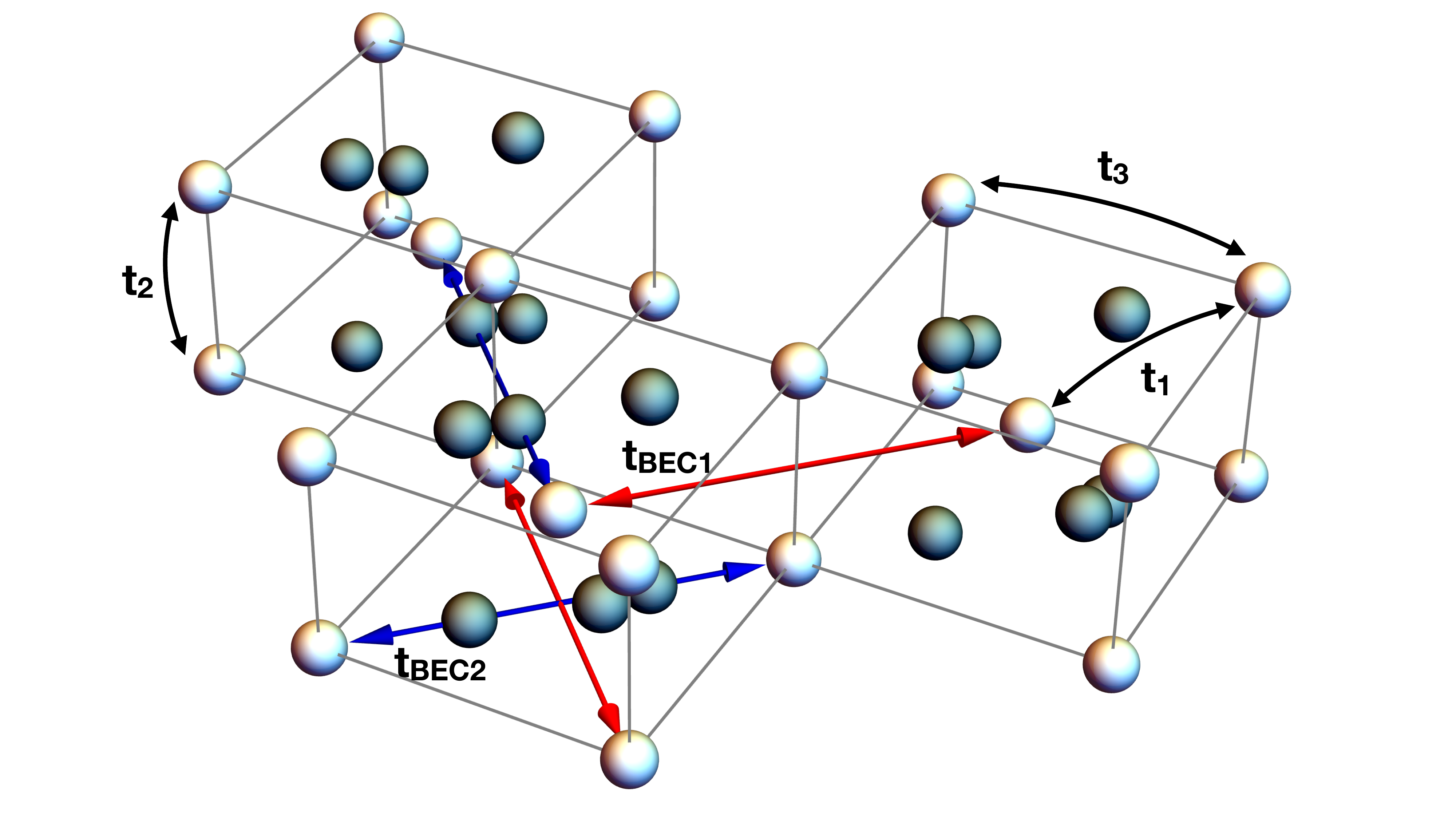}
    \caption{Illustration of the rutile structure discussed in the main text. The lattice formed by the magnetic ions (white spheres) is body-centred tetragonal. The decorating ions (gray spheres) break down some of the symmetries of that lattice. In particular, a $C_4$ symmetry is lost. For the nearest neighbor hopping $t_1$ as well as for $t_2$ and $t_3$ this loss of symmetry is irrelevant. But for the nearest neighbor hopping in the $[110]$ direction the absence of $C_4$ splits bonds at this shell into two inequivalent sets of bonds labelled in blue and red with their respective hopping integrals indicated. This is the shortest range hopping for which there is more than one color.}
    \label{fig:rutile}
\end{figure}

The three dimensional example we present is based on the structure of MnF$_2$ or RuO$_2$. Fig.~\ref{fig:rutile} shows three unit cells of this crystal where the white spheres  are the Mn sites and the black blobs the F sites. The former lead to a body-centered tetragonal crystal and the latter break the translation symmetry connecting the corner vertices and the cell center vertices. 
Beyond the simple translation symmetries of the crystal, there are two-fold rotation axes, $[001]$ and $[110]$. More importantly, there are non-symmorphic elements $-$ combinations of mirrors or rotations and fractional translations. In particular, there is a four-fold rotation symmetry around the $001$ axis accompanied by a translation connecting the corner and center vertices. It is this symmetry element that replaces the pure translation symmetry that would be there if the F ion decoration were not present. In figure~\ref{fig:rutile} we show the F sites alongside the Mn sites (even though the latter will be our primary concern) firstly to emphasise that the structure is connected to materials but most of all to give a visual guide to the lattice symmetries. Once again, the presence of the F ion ensures that the non-symmorphic four-fold symmetry is present and not the pure translation.

Having established the essential symmetries of the structure, we may now consider hopping on the lattice between Mn sites alone. The nearest neighbor hopping is between the corner vertices and the center vertices (labelled $t_1$ in Fig.~\ref{fig:rutile}). Each site is connected to eight others and the lattice symmetries map those bonds into one another. Therefore, for $s$-wave hopping, the tight-binding model for hopping between nearest neighbors has a single parameter. 

One may also show that the nearest neighbor hopping, $t_2$, along the $c$ direction has a single hopping parameter. So too does the nearest neighbor hopping in the 100/010 direction, $t_3$. The situation is different for hopping across the in-plane diagonal of the unit cell $(110)$ as indicated in Fig.~\ref{fig:rutile}. If the unit cell edge lengths are $a$ and $c$, these bonds have length $\sqrt{2}a$. The bonds that extend across this length split into two symmetry classes that cannot be mapped onto one another using the lattice symmetry operations. In the figure (Fig.~\ref{fig:rutile}) we have assigned colors blue and red to these inequivalent bonds. The root of this inequivalence is the absence of a pure four-fold rotation element among the space group elements. The result is a pair of hopping parameters (labelled $t_{\rm BEC1}$ and $t_{\rm BEC2}$).

The existence of such bond equivalence classes is central to this work. We find that they pervade the space groups and crystal structures associated to those space groups. 

\subsection{Colorings and bond equivalence classes}

In the above examples we saw that a given shell, specified by a fixed distance between neighboring sites, may have bonds that break up into classes that cannot be connected by lattice symmetries. These we call {\it bond equivalence classes} and to each set of bonds belonging to a class we assign a color exactly as in the examples discussed above. 

As we shall see in more detail through the course of this paper, the existence of more than one bond equivalence class at a given shell signals the breakdown of symmetry compared to the symmetry of the set of points forming the crystal structure. At first sight, one might expect an explosion of such bond equivalence classes with increasing hopping range because the number of group elements is fixed while the number of bonds within a spherical shell increases as $r^2$. However, at fixed shell distance, the number of bonds is of the order of the number of group elements and it is the distance between successive shells that shrinks with distance.

Here we illustrate this feature through the rutile example. For this case, one may address the symmetry of the bonds in a given shell. 
Under the space group symmetry elements, as all eight nearest neighbor bonds are equivalent, from the point of view of the tight-binding model for that shell, the lattice is effectively body-centred tetragonal with a single site in the primitive cell. This model is therefore more symmetric than the parent space group. One can see this also from the fact that the nearest neighbor hopping model has a four-fold rotational symmetry that is absent in the parent space group. 

The shortest hopping path that breaks the symmetry down to the parent crystal symmetry is the first shell with more than one bond equivalence class. This is the set of diagonal in-plane bonds described above. 

Having seen that the existence of multiple colors is tied to a breakdown of symmetry one is led to investigate whether this idea can be made more precise. In Section~\ref{sec:coloring} we carefully define the concept of bond equivalence classes and bond colors. We then show that the number of colors is exactly computable, in a broad class of cases, from an index connecting the symmetries of the set of points to the symmetry of the tight-binding model on the fixed shell. We can spell out exactly when this index is sufficient to capture the number of colors and why we cannot expect to find a similar index that computes the number of colors in all instances. We devote appendix~\ref{sec:exceptions} to explaining the manner in which the index result fails such that the number of colors is less than the index would predict. 

\subsection{Application to all space groups and Wyckoff positions}

One of the major results of this work is a systematic exploration of the symmetries of tight-binding models of finite range $s$-wave hopping. At the end of the paper, we present tables with the symmetry group of tight-binding models at the $n$th shell (where $n=1$ is nearest-neighbor hopping) for {\it all} space groups and {\it all} Wyckoff positions and for $n\leq 20$. 

Section~\ref{sec:general} describes how we determine the bond symmetries and provides a guide to the tables of enhanced symmetries resolved by shell number. There is much to be learned from this information. For a start, this considerable collection of data reveals that symmetry enhancement of bonds beyond that of the starting $-$ or parent $-$ space group is quite common. For a given parent, up to seven different crystal structures may be necessary to capture the symmetries of the shells. One finds that the translation symmetry may change or the point group or both. In the rutile example, the real space unit cell is smaller than that of the parent cell when one considers nearest neighbor bonds. But, in addition, there are additional symmorphic point group symmetries. Further observations of this nature including statistical properties of the symmetries can be found later on in Section~\ref{sec:general}. Among the more striking results are instances where the symmetry appears never to break down to the parent space group  such that arbitrarily long-range hopping models have enhanced symmetry. We identify candidates for this phenomenon through an empirical study out to large shell numbers $n\gg 20$ and we present a proof of this fact for a single instance.  

One problem that this investigation touches on is how to classify tight-binding models. One step in this direction hinted at by the symmetry information is to define the analog of a lattice complex for the bonds. These {\it bond complexes} are the possible space group and Wyckoff positions that can arise via symmetry enhancement of the bond-decorated lattice at a given shell number. They form a subset of all possible space groups and Wyckoff positions. The set of all tight-binding models are then naturally grouped into different bond complexes. This is further discussed in Section~\ref{sec:general}.

In general tight-binding models tend to involve hopping parameters that fall off with distance. Therefore it is of interest to investigate the symmetries obtained by combining bonds for all shells up to a fixed cutoff. These cumulative results are all presented in Section~\ref{sec:general} together with a discussion of their properties. 

Section~\ref{ssec:beyond} is a description of some features of the bonds that go beyond their symmetry. These are potentially important if one wishes to understand the band structures originating from the bonds. 

Section~\ref{ssec:bands} is a partial investigation of band structures for different shells where there are enhanced symmetries. We remark on the role of topological quantum chemistry when there are symmetry enhancements. 

Finally, we comment on applications of the work here to altermagnetism.

\section{Crystals and crystalline symmetries}
\label{sec:crystal}

Here we give a short overview of the main concepts of crystal symmetries that underlie this work. Crystalline structures of solids are common in nature and are routinely characterized using X-ray or neutron diffraction. Crystals are invariant under discrete real space translations of the form $\sum_{a=1,2,3} n_a \mathbf{R}_a$ where $\mathbf{R}_a$ are the so-called {\it primitive translation vectors}. These translations form an infinite (Abelian) group $\mathbf{T}$ and the set of points obtained by acting with these translations on a single point is called a {\it lattice}.  There are fourteen distinct {\it Bravais lattices} in three dimensions classified according to the symmetry of the primitive cell which are further divided into seven crystal classes e.g. the cubic crystal class subsuming simple cubic, bcc and fcc Bravais lattices. 

Real crystals are lattices decorated with some number of atoms $-$ the basis $-$ within each primitive cell and their symmetries form so-called {\it space groups}. There are $230$ space groups in three dimensions and each one has a collection of elements
\beq
\bigcup_\alpha \left\{ S_\alpha \vert \ \mathbf{t}_\alpha \right\} \mathbf{T} .
\eeq
Operations with trivial translation part $\left\{ S_\alpha \vert \ \boldsymbol{0} \right\} $ can be rotations, mirrors, inversion as well as roto-reflections and roto-inversions  belonging to one of the $32$ points groups that are compatible with a tiling of three dimensional space. There can also be elements where a rotation is combined with a fractional translation, $\mathbf{t}_\alpha$ (screw operations) or where a reflection is combined with a translation (glide operations). The $73$ space groups with neither screws nor glides are called {\it symmorphic} space groups. The remainder which form a majority are non-symmorphic. 

Let us consider more closely the positions within the primitive cell. A given point may be left invariant by some subset of the point group elements of the space group. This set of elements is called the {\it site symmetry group} (SSG). Every point has some SSG (which may be trivial in which case it is called a general position) and the positions whose SSGs are conjugate to one another for a given space group form a {\it Wyckoff position}. The elements of the space group that do not belong to the SSG shift the position and the number of points in the primitive cell obtained by acting in this way for a given Wyckoff position is called the multiplicity of that Wyckoff position. Wyckoff positions, by convention, carry a label which is the multiplicity followed by a letter. Across all space groups there are $1731$ Wyckoff positions. In a real crystal, an atom or ion is assigned to a particular Wyckoff position.  

As explained previously, one of the central ideas in this work is that the generating space group of a tight-binding model cannot necessarily be inferred from the tight-binding model itself. A similar idea arises in the case of sets of points in space. The set of points generated by a space group without reference to that space group is called a {\it point configuration}. The maximal space group that generates a point configuration is called the {\it eigensymmetry group} of the point configuration. A further useful piece of terminology is that of characteristic and non-characteristic orbits \cite{wondratschek1976extraordinary,wondratschek1980crystallographic,engel2015non}. The crystallographic orbit is the set of points generated from a given point under the action of a space group. If the eigensymmetry group of the resulting point configuration matches that of the generating space group then the orbit is said to be {\it characteristic} and {\it non-characteristic} otherwise. 

Now if we take two Wyckoff positions from two different space groups {\it belonging to the same crystal class} and form their point configurations we say that these point configurations belong to the same {\it lattice complex} if they can be mapped into one another by some combination of rotations, translations or anisotropic scale transformations (the latter of which must preserve the crystal class) \cite{wondratschek1980crystallographic,brock2016international}. One simple example is that multiple Wyckoff positions of certain space groups generate point configurations that are primitive cubic lattices. The idea of a lattice complex unifies these point configurations across different space groups and Wyckoff positions. It turns out that there are $402$ lattice complexes. These are used to organize the results into bond complexes. Therefore a complete list of lattice complexes can be found at the end of this paper. 

Some of our central results relate to the number of bond equivalence classes. To understand these results in detail it is important to introduce some ideas of group-subgroup relations for space groups and, especially, the concept of a crystallographic index. 

A subgroup $H$ of a group $G$ has a subset of the elements of $G$ while still obeying the group relations. For space groups, it has proved useful to distinguish two types of subgroup: {\it translationengleiche} and {\it klassengleiche} hereafter t-subgroup and k-subgroup respectively. A t-subgroup subgroup $H$ of $G$ has the same pure translation elements $T(G)$ as $G$, $T(G)=T(H)$ but fewer cosets $\vert H/T(H)\vert<\vert G/T(G)\vert$ meaning that the order of the $H$ coset is less than that of $G$. A k-subgroup instead retains the point group of $G$, $\vert H/T(H)\vert = \vert G/T(G)\vert$, but has less translation symmetry $T(H)<T(G)$ bearing in mind that the number of translation elements is infinite. 

Of course, there are subgroups that are neither of the k-subgroup nor t-subgroup varieties. But, thanks to Hermann \cite{delaFlor2023}, we know the following. Suppose we have space groups $G$ and $H$ where $H$ is a subgroup of $G$. Then there exists a subgroup $M$ of $G$: $H\leq M \leq G$ such that $M$ is a t-subgroup of $G$ and such that $H$ is a k-subgroup of $M$. 

Now it is useful to introduce a pair of indices. One is integer $[i_k]$ which is the ratio of the number of atoms in the primitive cells $N(H)/N(G)$. The second is integer $[i_t]$ which is the ratio of the orders of the point groups $\vert G \vert / \vert H \vert$. A k-subgroup is connected to its supergroup via index $[i_t]=1$ and nontrivial $[i_k]$ index. A t-subgroup has $[i_k]=1$ and nontrivial $[i_t]$ index. Finally, a general subgroup also carries an index $[i]=[i_t]\times [i_k]$. We make use of these indices in the next section. 

\section{Detailed Example}
\label{sec:tbmodelexample}

In Section~\ref{sec:overview} we illustrated the ideas of symmetry enhancement and bond equivalence classes through two examples $-$ one in two dimensions and one in three dimensions. Our goal is to explore symmetry enhancement across all space groups and Wyckoff positions. To this end, this section enters into more technical details of the computation of the tight-binding model symmetries. At the same time, we shall see one  instance of how symmetries may change when the shell number changes. We focus, here, on space group $P4$ (\#75) and Wyckoff position 2$c$ for different shell numbers. This parent crystal exhibits a rich hierarchy of symmetry enhancements at different shell numbers, involving breakdown of translation symmetries, point-group symmetries, or both.

The primitive cell is tetragonal with a two site basis with coordinates $(\frac{1}{2},0,z)$ and $(0,\frac{1}{2},z)$. The {\it eigensymmetry group } (the maximal space group for the point configuration generated by the parent data) for this case is $P4/mmm$ (\#123). As this will be important in the following, we consider the connection between this and the parent $P4$. The elements of $P4mmm$ are listed in Table~\ref{tab:P4mmm} with the subset belonging to $P4$ highlighted in bold. In fact, the standard cell of $P4$ is related to that of $P4/mmm$ by the transformation
\begin{equation}
 T= \left( \begin{array}{ccc|c}
   1 & 1 & 0 & 1 \\
   -1 & 1 & 0 & 0 \\
   0 & 0 & 1 & -z\\
\end{array}\right)
\end{equation}
which is not an isometry as it shrinks the unit cell by a factor of two. In particular, the basis of $P4/mmm$ has two lattice vectors which are non-trivial (not lattice translations) in the basis of $P4$. Indeed they are fractional translations: $(\tfrac{1}{2} \ \tfrac{1}{2} \ 0)$ and $(\tfrac{1}{2} \ -\tfrac{1}{2} \ 0)$. Similarly to the rutile case for the nearest neighbor shell, this results in a smaller unit cell, as the lattice translations now relate the two distinct sites at Wyckoff position 2$c$. Thus, the lattice goes from having a Wyckoff position of multiplicity $2$ ($2$ sites per unit cell) to multiplicity $1$ ($1$ site per unit cell). In other words, the Wyckoff position $2c$ of $P4$ maps to Wyckoff position $1c$ of $P4/mmm$ corresponding to the same point configuration. 
\begingroup
\begin{widetext}
\begin{center}
    \setlength{\tabcolsep}{7pt} 
\renewcommand{\arraystretch}{1.5} 
\begin{table}[h]
    \centering
    \begin{tabular}{cccccc}
    $\boldsymbol{\{1|0\}}$ &  $\boldsymbol{\{4^{\pm}_{001}|0\}}$  & $\boldsymbol{\{2_{001}|0\}}$ & $\{2_{010}|0 \ 0\  2z\}$ &  $\{2_{100}|0 \ 0 \ 2z\}$ & $\{2_{1\pm10}|0 \ 0 \ 2z\}$\\
     $\{-1|0 \ 0 \ 2z\}$ &  $\{-4^{\pm}_{001}|0 \ 0 \ 2z\}$  & $\{m_{001}|0 \ 0 \ 2z\}$ & $\{m_{010}|0\}$ &  $\{m_{100}|0\}$ & $\{m_{1\pm10}|0\}$\\
     & & $\{1|\frac{1}{2} \ \frac{1}{2} \ 0\}$ & $\{1|\frac{1}{2} \ -\frac{1}{2} \ 0\}$ & &
    \end{tabular}
    \caption{List of elements of $P4/mmm$ expressed in the standard reference of $P4$, up to trivial lattice translations. The elements of $P4$ are highlighted in bold.}
    \label{tab:P4mmm}
\end{table}
\end{center}
\end{widetext}
\endgroup
Now we investigate the symmetries of the hopping models for a given shell number. As the highest symmetry is the eigensymmetry group, it is sufficient to check which symmetries in Table~\ref{tab:P4mmm} are respected by any given hopping model. For shell number one, the hopping model consists of decoupled layers with four bonds in each plane per unit cell. These are related by $\{4^\pm_{001}|0\}$ and therefore constitute a single equivalence class. For the second shell, there are two vertical bonds per cell so the hopping is now quasi-1D and again the bonds are related by $\{4^\pm_{001}|0\}$ and form a single equivalence class. In fact, precisely because there is a single equivalence class for both shells, the eigensymmetry group is the symmetry group of these subdimensional hopping models.

\begin{figure}[h!]
    \centering
    \includegraphics[width=4cm]{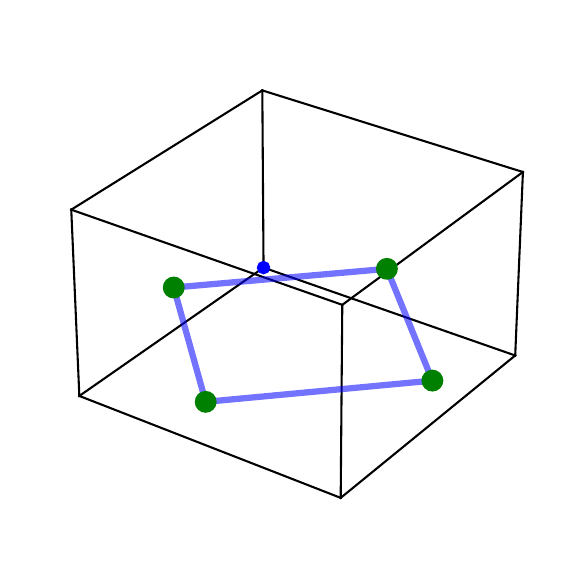}
    \includegraphics[width=4cm]{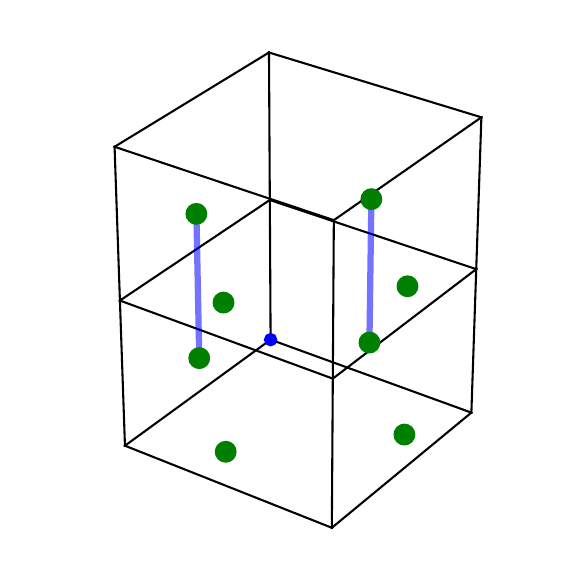}
    \caption{First and second shell bonds for space group $P4$ and Wyckoff position $2c$, which each form a single equivalence class.}
    \label{fig:1plus2}
\end{figure}

\begin{figure}[h!]
    \centering
    \centerline{\includegraphics[width=4.4cm]{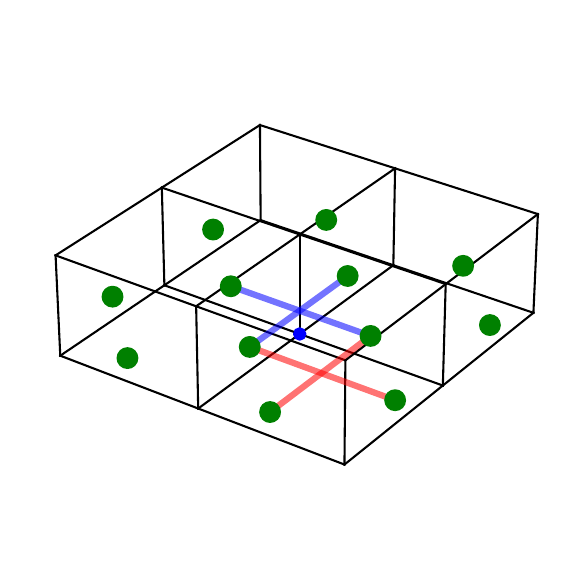}
    \includegraphics[width=4.4cm]{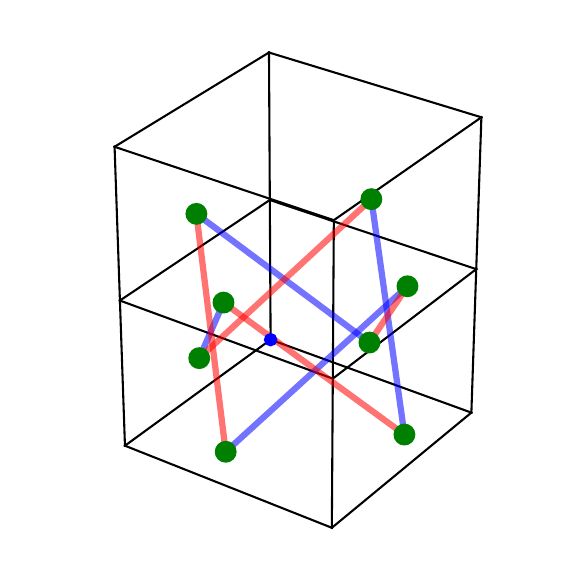}}
    \caption{Third and fourth shell bonds for space group $P4$ and Wyckoff position $2c$.}
    \label{fig:3plus4}
\end{figure}

The situation becomes more interesting at the third shell. As for the first shell, the hopping is two dimensional but now with four bonds divided into two equivalence classes as illustrated in Fig.~\ref{fig:3plus4}. Representative bonds are
\begin{equation}
   \Bigg\{ \begin{pmatrix}
        \tfrac{1}{2} & 0 & z\\
        \tfrac{3}{2} & 0 & z
    \end{pmatrix},\begin{pmatrix}
        0 & \tfrac{1}{2} & z\\
        0 & \tfrac{3}{2} & z
    \end{pmatrix}\Bigg\}, \ \Bigg\{\begin{pmatrix}
        \tfrac{1}{2} & 0 & z\\
        \tfrac{1}{2} & 1 & z
    \end{pmatrix}, \begin{pmatrix}
        0 & \tfrac{1}{2} & z\\
        1 & \tfrac{1}{2} & z
    \end{pmatrix}\Bigg\}
\end{equation}
where the two equivalence classes are in separate curly brackets. The central consequence of the two inequivalent sets of bonds is the breaking of the fractional translation symmetry of $P4/mmm$. All other symmetries of $P4/mmm$ are still symmetries of the bonds, however, so the overall hopping model still belongs to $P4/mmm$ but now with Wyckoff position $2f$ instead of $1c$. 

For the fourth shell there are eight bonds divided into two equivalence classes as shown in Fig.~\ref{fig:3plus4}. 
\begin{align}
   &\Bigg\{ \begin{pmatrix}
        \tfrac{1}{2} & 0 & z\\
        0 & \tfrac{1}{2} & z+1
    \end{pmatrix},\begin{pmatrix}
        \tfrac{1}{2} & 0 & z+1\\
        1 & \tfrac{1}{2} & z
    \end{pmatrix},\begin{pmatrix}
        1 & \tfrac{1}{2} & z+1\\
        \tfrac{1}{2} & 1 & z
    \end{pmatrix},\begin{pmatrix}
        \tfrac{1}{2} & 0 & z+1\\
        1 & \tfrac{1}{2} & z
    \end{pmatrix}\Bigg\}, \\
    &\Bigg\{ \begin{pmatrix}
        \tfrac{1}{2} & 1 & z\\
        1 & \tfrac{1}{2} & z+1
    \end{pmatrix},\begin{pmatrix}
        \tfrac{1}{2} & 1 & z+1\\
        1 & \tfrac{1}{2} & z
    \end{pmatrix},\begin{pmatrix}
        0 & \tfrac{1}{2} & z+1\\
        \tfrac{1}{2} & 1 & z
    \end{pmatrix},\begin{pmatrix}
        \tfrac{1}{2} & 0 & z+1\\
        0 & \tfrac{1}{2} & z
    \end{pmatrix}\Bigg\}
\end{align}
This is the first shell that fully couples the sites in three dimensions. Nevertheless the symmetry is higher than the parent symmetry and lower than the eigensymmetry. This time, on top of breaking the lattice translation symmetries $\{1\vert \frac{1}{2} \ \pm\frac{1}{2} \ 0\}$, the (non-symmorphic) inversion symmetry $\{-1|0 \ 0 \ 2z\}$ is also broken, and thus the entire second row of elements in $P4/mmm$ as listed in \cref{tab:P4mmm}. Since both $\{1\vert \frac{1}{2} \ \pm\frac{1}{2} \ 0\}$ and $\{-1 \vert 0 \ 0 \ 2z\}$ toggle between the two equivalence classes, their composition must necessarily be a symmetry. Therefore, the list of symmetries up to trivial lattice translations is given by the first row of \cref{tab:P4mmm} and the second row composed with the third row. Overall, the space group is $P4/nbm$ (\# 125), and the Wyckoff position is 2$c$.

One may continue to explore symmetries of the bonds shell-by-shell in this manner. Up to the 20th shell one finds six enhanced symmetry groups and associated Wyckoff positions for the parent group $P4$ and Wyckoff position $2c$ bounded by the parent symmetry group and the eigensymmetry group. This example also highlights the potential for hopping models at a given shell number to be subdimensional $-$ on decoupled chains or planes. Later examples will also reveal the potential for hopping to break up into zero dimensional clusters as well as decoupled sublattices.

\section{Colorings, bond equivalence classes and color index}
\label{sec:coloring}
In this section, we provide a rigorous introduction to the concepts of bond complexes, bond equivalence classes/bond colorings on lattice graphs, and their counting based on group-subgroup indices and the orbit-stabilizer theorem. We thus employ a group-theoretic approach based on the notion of group actions on finite sets. 

\subsection{Colors and Bond Equivalence Classes Defined}

Suppose we are given a parent space group $G$ and Wyckoff position $\mathcal{W}$, which together form the {\it parent crystal data} $(G,\mathcal{W})$. By $\mathcal{W}$ we denote the set of all sites $w=(x,y,z)$ in the lattice corresponding to the given Wyckoff position. We denote a bond on this lattice as an unordered pair $\{w,w'\}$ indicating its end-points, which must be sites of the lattice. At any given shell order $n$, we can then construct the set $\mathcal{B}_n$ of bonds in that shell by enumerating all unordered pairs of sites $\{w,w'\}$ that are separated by a distance $d_n$ dictated by the shell order:
\begin{equation}
    \mathcal{B}_n = \{\{w,w'\} \ | \ w,w' \in \mathcal{W} \text{ such that } |w-w'|=d_n\}
\end{equation}

Having found $\mathcal{B}_n$, one may then consider the action of $G$ on this set. In particular, let us define an equivalence relation $\sim$ between two bonds indicating if they can be transformed into each other under the action of some element of $G$:
\begin{align}
     \{w_1,w_2\} \sim \{w_1',w_2'\} \iff \forall i=1,2, &\exists g \in G: w_i=gw_j' \\ \nonumber
     &\text{ for some $j=1,2$.}
\end{align}
Evidently each bond is equivalent to itself as the identity operation simply leaves the bond invariant and there may be operations that merely exchange the sites. 
Then, the distinct orbits of the bonds in $\mathcal{B}_n$ under the action of $G$ generate all the bond equivalence classes:
\begin{equation}
    [\{w_1,w_2\}] = \{\{w_1',w_2'\}\ | \ \{w_1,w_2\} \sim \{w_1',w_2'\}\} = G\{w_1,w_2\}.
\end{equation}
Let us define the set of bond equivalence classes 
\begin{align}
    \mathcal{B}_n/G&=\{[\{w,w'\}] \ | \ \forall \{w,w'\} \in \mathcal{B}_n \} \nonumber \\
    &= \{G\{w,w'\} \ | \ \forall \{w,w'\} \in \mathcal{B}_n \}.
\end{align}
Having multiple bond equivalence classes thus corresponds to having sets of inequivalent bonds, i.e. bonds which cannot be mapped into each other by a symmetry of the parent space group. In a tight-binding model, the hopping coefficients on these bonds are not related to each other. Visually, we can imagine assigning a color to each bond equivalence class and decorating the lattice by coloring each bond according to which equivalence class they belong to. This decoration can lead to a breakdown of symmetries as we shall now explore. 

By the breakdown of symmetries we mean that the symmetries are reduced relative to the eigensymmetry group corresponding to the symmetries of the points in space generated by the space group and Wyckoff position. One finds that certain shells with the full eigensymmetry occasionally have more than one bond equivalence class. We saw this even in the kagome example: if the nearest neighbor distance is $a$, there are two equivalence classes (two colors) among the bonds at distance $2a$ when the full eigensymmetry is considered (in Fig.~\ref{fig:kagome} these are (i) the bonds across the hexagon and (ii) the merged blue and red bonds as, for the eigensymmetry group, the blue and red bonds belong to the same bond equivalence class). 
In general, for a given shell number $n$ such that the hopping model has the eigensymmetry we denote the number of bond equivalence classes by $b_{\rm ES}^{(n)}$. In the kagome case we have $b_{\rm ES}^{(3)}=2$.  Whether or not $b_{\rm ES}^{(n)}>1$ is dependent on the details of the lattice and we should not expect to be able to compute it from a simple index. 

\subsection{Index Theorem for the Number of Colors}

For simplicity we consider first $b_{\rm ES}^{(n)}=1$. We also take $[i_k]=1$ defined to be the translation index connecting the eigensymmetry group and the symmetry group of the tight-binding model. With these conditions, that we relax later, we consider a general shell that may or may not have the symmetries of the parent group or the eigensymmetry group. At this shell there are $b^{(n)}$ bond equivalence classes. In general we are interested  to know which elements of the eigensymmetry group $E=\text{Eig}_\mathcal{W}(G)$ of $(G,\mathcal{W})$ are still present taking the bond equivalence classes into account. The resulting group $H$ is a subgroup of $E$, and can be defined as the stabilizer of the bond equivalence classes:
\begin{align}
    H = \text{Stab}_E(\mathcal{B}_n/G) = \{g \in E \ |\  & g\{w_1,w_2\}\in [\{w_1,w_2\}] \nonumber \\ & \forall \{w_1,w_2\} \in \mathcal{W}\}.
\end{align}

Now we find that, when the translation group-subgroup index $[i_k]=1$, the number of colors, or bond equivalence classes, can be computed from the orbit-stabilizer theorem \cite{bradley2009mathematical}. In particular, 
\begin{equation}
  b^{(n)} = |\mathcal{B}_n / G| = |E:H| \equiv [i_t]
\label{eq:orbitstabilizer}
\end{equation}
which relates the number of bond equivalence classes to the group-subgroup index of the eigensymmetry group and enhanced subgroup. Or, in words:
\begin{quote}
{
\bf Index
} When the number of colors for the eigensymmetry equals one and when the translation index is one connecting the eigensymmetry group to the crystal at the chosen shell number, the number of colors equals the total group-subgroup index for the respective point groups of the eigensymmetry group and enhanced subgroup at a given shell number. 
\end{quote}

We have computed the index and number of colors for all space groups and Wyckoff positions and for shells up to $n\leq 20$ in our convention. Out of all these cases, roughly $80\%$ satisfy the conditions $b_{\rm ES}^{(n)}=1$ and $[i_k]=1$ and, for all such cases, Eq.~\ref{eq:orbitstabilizer} correctly computes the number of colors. 

We now relax the conditions. We first consider cases where $b_{\rm ES}^{(n)}=1$ and $[i_k]\neq 1$. In other words, we consider cases where the volume of the unit cell changes between the lattice with eigensymmetry and the tight-binding model. There are $3580$ such cases for $n\leq 20$. We conjectured that the number of colors is equal to the total index $[i_k][i_t]$. We find that this works in $3078$ of the cases while $1/2[i_k][i_t]$ computes the colors in a further $344$ cases leaving $239$ cases where both of these guesses fail.

We then consider $b_{\rm ES}^{(n)}> 1$ and $[i_k]= 1$. Now there are only $936$ cases. Here we explore the possibility that the bond equivalence classes with eigensymmetry both split into further colors when the symmetry is reduced. If this were the case we would expect 
\begin{equation}
  b^{(n)} \overset{?}{=} b^{(n)}_{\rm ES} |\mathcal{B}_n / G| = b^{(n)}_{\rm ES} |E:H| = b^{(n)}_{\rm ES}[i_t].
\label{eq:orbitstabilizer2}
\end{equation}
We find that this formula is correct in about $600$ cases. Therefore, eigensymmetry bond equivalence classes do not behave in the same way when the symmetry is reduced. 

Finally we consider the few remaining cases: those with $b_{\rm ES}^{(n)}> 1$ and $[i_k]> 1$. There are $204$ cases. We conjecture ({\bf Index$^*$})
\begin{equation}
  b^{(n)} \overset{?}{=}  b^{(n)}_{\rm ES}[i_t][i_k]
\label{eq:orbitstabilizer3}
\end{equation} 
finding that this works in $44$ instances.

In addition to a breakdown of symmetries, there may also be a shrinking of the unit cell, as we saw in the example of P4(2$c$). This means that the multiplicity of the Wyckoff position of $H$ $\mathcal{W}$ may differ from that of the eigensymmetry group. While the set of sites will be the same, the reference system they will be expressed in is generally different for $H$ and $G$, so we obtain a new set of site coordinates $\mathcal{W}'$ in addition to $H$. The enhanced crystal data is thus $(H,\mathcal{W}')$.

The details of how the bond equivalence classes $\mathcal{B}_n/G$, the enhanced space group $H$ and Wyckoff position $\mathcal{W}'$ are identified using the tools in the Bilbao crystallographic server \cite{bcs1,bcs2,aroyo2006bilbao} are given in Appendix~\ref{app:algorithm}. 

\subsection{Defining a bond complex}
The algorithm described in Appendix~\ref{app:algorithm} provides a map $\phi_n$ from $(G,\mathcal{W})$ to $(H,\mathcal{W}')$ at any shell order $n$. It is clear that the domain and image of $\phi_n$ must be the set of 230 space groups and 1731 Wyckoff positions. However, there are additional constraints limiting what $(G,\mathcal{W})$ can enhance to. Indeed, we note that $G$ and $H$ are both subgroups of $E$, and since $E$ is a minimal supergroup of both, it follows that  $(G,\mathcal{W})$ and $(H,\mathcal{W})$ must have the same eigensymmetry group $E$. In other words, $(G,\mathcal{W})$ can only enhance to a space group and Wyckoff position within its lattice complex LC$(G,\mathcal{W})$. We can then ask ourselves, within a given lattice complex, which groups and Wyckoff positions enhance to a particular $(H,\mathcal{W}')$? This leads us to the notion of a bond complex: 
\begin{align}
\text{BC}(H,\mathcal{W}') & = \{(G,\mathcal{W}) \in LC(H,\mathcal{W}') \ |  
\nonumber \\ & \exists n \in \mathbb{N}, \ \phi_n(G,\mathcal{W}) = (H,\mathcal{W}')\}
\end{align}
The analogy between lattice complexes and bond complexes is evident. A lattice complex answers the question: which space group and Wyckoff position generate a point configuration with a specific eigensymmetry group and corresponding Wyckoff position? Analogously, the bond complex answers the question: which space group, Wyckoff position and shell (which decorates the lattice with bond colorings) can enhance to a particular space group and Wyckoff position?

\subsection{Lattice Complexes and Group-Subgroup Relations}

We have already observed that the members of a hierarchy of hopping models for different shell number belong to a common lattice complex with the parent crystal. But it is not true that all space groups and Wyckoff positions for a given lattice complex appear within a hierarchy. It should perhaps not be surprising that the set of hopping models forms a restricted class among the space groups and Wyckoff positions belonging to a lattice complex. One important question is whether there is a simple symmetry criterion that establishes the relationship between the members of a hierarchy.

A nontrivial k-subgroup $H$ of $G$ may be isomorphic to $G$. For example, one may imagine doubling the unit cell in some direction retaining all other symmetries. The number of such isomorphic subgroups is evidently infinite whereas the number of non-isomorphic k-subgroups is finite. 

 Let us now inspect the groups appearing in a hierarchy using tables of group-subgroup relations for maximal subgroups of both k-type and t-type. We know that the eigensymmetry is the highest symmetry and the other groups appearing for different shell numbers must be subgroups.
 
Our first example is the hierarchy of length three belonging to $\# 230 (16a)$ that includes  $\# 229 (2a)$ and  $\# 223 (16a)$. $\# 230$ has the lowest symmetry and is the maximal k-subgroup of $\# 223$ which is itself a maximal k-subgroup of $\# 229$. So the web of connections here only involve changes to the translation symmetry.  

Another example is the hierarchy of length five coming from $\# 195 (6i)$ that includes groups $\# 215$, $\# 200$, $\# 207$, $\# 221$. These are all cubic groups and the web of connections is through t-subgroups alone. The highest symmetry group among these is $\# 221$ which is one step removed from the three t-subgroups $\# 200$, $\# 207$ and $\# 215$. And then $\# 195$ is both one step lower than $\# 207$ and one step lower than $\# 200$ in a graph of t-subgroups. 

We have seen an example of a hierarchy through k-subgroups and another through t-subgroups. Now we consider an example with both k- and t-subgroups. This is based on $\# 209 (24d)$ and is a hierarchy of length five with, in addition, $\# 207$, $\# 221$, $\# 225$, $\# 226$. We start with a group of highest symmetry here: $\# 221$ which is symmorphic. In a graph of k-subgroups $\# 221$ is on the same level as $\# 225$ which is, however, non-symmorphic. Then $\# 226$ is a maximal k-subgroup of $\# 221$. Then $\# 207$ is a t-subgroup of $\# 221$ and finally $\# 209$ is a k-subgroup that includes fractional translations in addition to the point group elements of symmorphic $\# 207$. 

In summary, any group in a hierarchy is either the group of highest symmetry or is a maximal subgroup of another group in the hierarchy. The type of subgroup may be of k-type or t-type and there are hierarchies that are exclusively through k-subgroups, exclusively through t-subgroups or mixed.

\section{Exploring Symmetry Enhancement among the Space Groups}
\label{sec:general}
Having covered the theoretical underpinnings of the concepts of lattice complexes, shell anomalies and bond complexes in a rigorous manner, we now present the results obtained from the classification of bond complexes (up to the 20th shell). We begin by providing a guide to the tables listing different instances of the shell anomaly, and their regrouping into tables listing bond complexes. 

\subsection{Guide to the Symmetry Enhancement Tables }
\label{ssec:guide}
The most detailed form of the results in this paper are presented in Appendices~\ref{app:tab_enh_shell} and~\ref{app:bond_complexes} as tables. We thus provide a short guide to these tables so that they may be referenced with ease. 

The tables in Appendix~\ref{app:tab_enh_shell}, which we call \enquote{Hierarchy tables}, show which space groups and Wyckoff positions an initial parent crystal data can enhance to within the first 20 shells. They thus list the {\it hierarchy} of shell anomalies for a given parent crystal data. The first two rows, titled \enquote{Space group} and \enquote{Wyckoff} respectively, identify the parent space group and Wyckoff position. The third column, titled \enquote{Enhanced Space Group}, lists the enhanced crystal data that the parent space group and Wyckoff position will enhance to within the first 20 shells. We denote this list as the parent crystal data's hierarchy. This will consist of a list of space groups with the Wyckoff position in parenthesis. However, if the parent crystal data's eigensymmetry group is the same, then there is no possibility for enhancement regardless of shell number, and thus we simply restate the original parent data followed by \enquote{for all shells}. We further note that the tables have been partitioned into different crystal classes to ease their browsing. For example, if one wants to view the hierarchy of space group 75 Wyckoff position 2c, then one must go to Table VII (corresponding to the tetragonal crystal class) and in the second row one finds the hierarchy listed as \enquote{85(2$a$), 83(1$c$), 123(2$f$), 127(2$d$), 123(1$c$), 125(2$c$)}. Similarly, for space group 76, Wyckoff position 4a, the hierarchy is listed as \enquote{76(4$a$) for all shells} because the eigensymmetry group is not enhanced from the parent space group.

The tables in Appendix~\ref{app:bond_complexes}, which we call \enquote{Bond complex tables}, reorganize the data in Appendix~\ref{app:tab_enh_shell} into different bond complexes. Since each bond complex is bounded to be within the parent lattice complex, we organize the bond complex tables according to which lattice complex they belong to; this is denoted at the top of each table in bold. The first column of the table then gives the label of the bond complex, while the second column lists the parent crystal data (space group, Wyckoff position in round brackets and shell number in square brackets) which enhances to that bond complex label. For example, if we want to view the bond complex of 85(2$a$), then we must go to the bond complex table corresponding to the lattice complex which 85(2$a$) belongs to, namely 123(1$a$). We then see that the bond complex of 85(2$a$) is listed as \enquote{75(2$c$)[12, 16], 85(2$a$)[12, 16]}. So 75(2$c$) enhances to 85($2a$) at the 12th and 16th shell, and likewise for 85(2$a$). 

We note that one important difference between Appendix \ref{app:tab_enh_shell} and \ref{app:bond_complexes} is that the latter provides shell-resolved data, while the former does not. So if one is looking for shell-resolved data then the bond complex table should be referenced. 

\begin{widetext}
\centering
\begin{table*}
\begin{tabular}{| c | c | c | c | c | c |} 
 \hline
  Crystal class  &  \multicolumn{1}{|p{1.5cm}|}{\centering Number \\ of  WPs} & \multicolumn{1}{|p{3cm}|}{\centering Percentage of WPs \\ with matching \\ eigensymmetry}  &  \multicolumn{1}{|p{3cm}|}{\centering Percentage of WPs \\ with no enhanced symmetry \\ up to the 20th shell}  & \multicolumn{1}{|p{3cm}|}{\centering Percentage of WPs \\ with  multiplicity one}  & \multicolumn{1}{|p{3cm}|}{\centering Percentage of WPs \\ which don't break down \\ up to 20th shell} \\
 \hline
 Triclinic & 10 & 90.0 & 90.0 & 90.0 & 10.0 \\  
 Monoclinic & 65 & 47.7 & 47.7 & 21.54 & 13.8 \\  
 Orthorhombic & 434 & 30.88 & 30.88 & 4.61 & 16.6 \\ 
 Tetragonal & 562 & 26.69 & 26.69 & 4.98 & 36.3 \\ 
 Trigonal & 132 & 32.58 & 32.58 & 15.91 & 28.8 \\  
 Hexagonal & 220 & 37.73 & 37.73 & 10.0 & 27.7 \\ 
 Cubic & 308 & 44.16 & 44.16 & 3.25 & 8.1 \\ 
 \hline
\end{tabular}
 \caption{Table giving data related to symmetry enhancement averaged over space groups and Wyckoff positions for each crystal class.}
\label{tab:overview}
\end{table*}
\end{widetext}

\subsection{A Survey of the Landscape}
\label{ssec:landscape}

Having calculated the symmetries of hopping models at different shell number for the different space groups and Wyckoff positions we look at the gross, statistical features of the results. Table~\ref{tab:overview} gives an overview of some such preliminary observations. The first point to note is that, of the seven crystal classes, some are much more prolific than others at producing Wyckoff positions. The tetragonal class has the most Wyckoff positions, $562$, and the largest number of groups of all the classes followed by the orthorhombic case with $434$. A necessary condition for the hopping model to have higher symmetry than the parent space group is that the eigensymmetry group be of higher symmetry. Of the ten triclinic Wyckoff positions (divided among two space groups) all but one has its eigensymmetry group equal to the parent space group. For the other crystal classes the proportion of Wyckoff positions admitting enhanced symmetries is significantly higher $-$ for example about three-quarters of the tetragonal Wyckoff positions have eigensymmetry of higher symmetry than the parent group. Inspection of the first two columns of Table~\ref{tab:overview} leads us to expect the richest sets of enhanced symmetries among the tetragonal and orthorhombic cases. But, for all but the triclinic class, we observe that the number of cases satisfying the necessary condition for symmetry enhancement is large so we should not be surprised to see, on the basis of these data, that the cases of symmetry enhancement are very numerous. 

The third column of Table~\ref{tab:overview} is also informative. The numbers in this column match those in the second column. In other words, it appears that the set of cases (space group and Wyckoff position) where no symmetry enhancement is possible matches the set of cases where it does not occur. In other words, if the eigensymmetry group is of higher symmetry than the parent data then symmetry enhancement does take place at least at some shell number. Symmetry enhancement is also bounded by the fact that it can only take place within a given crystal class. 

 One simple measure of the degree of which symmetry may be enhanced is the number of different space groups and Wyckoff positions that arise for a fixed choice of parent crystal data and as the shell number is varied. We first concentrate on the data where hopping takes place at a given shell number $n$. Below we consider the cases of symmetry enhancement where hopping takes place between shells up to and including $n$. The guide to the tables (Section~\ref{ssec:guide}) introduced possible hierarchies of symmetries. As another example, inspection of the table for orthorhombic space group $45$ and Wyckoff position $4b$ shows that three different cases may appear for different shell numbers: $72(4d)$, $72(4b)$, $65(2b)$ where the space group $72$ appears twice with different Wyckoff positions with the {\it same} multiplicity. Inspection of the table in Appendix D for the triclinic crystal class where only one of the ten Wyckoff positions may be enhanced shows that indeed that the only allowed case of enhancement (SG $\#1 (1a)$)) is enhanced at some shell numbers to SG $\# 2$. Peculiarly hopping models at any shell number (at least up to $n=20$) for SG $\#1(1a)$ are enhanced in symmetry. We explore the extent of this phenomenon below. Turning to the monoclinic class, we see that it is significantly richer than the triclinic class containing many hierarchies of length 2. 
 
\begin{figure*}[h!]
    \centering
    \includegraphics[width=2\columnwidth]{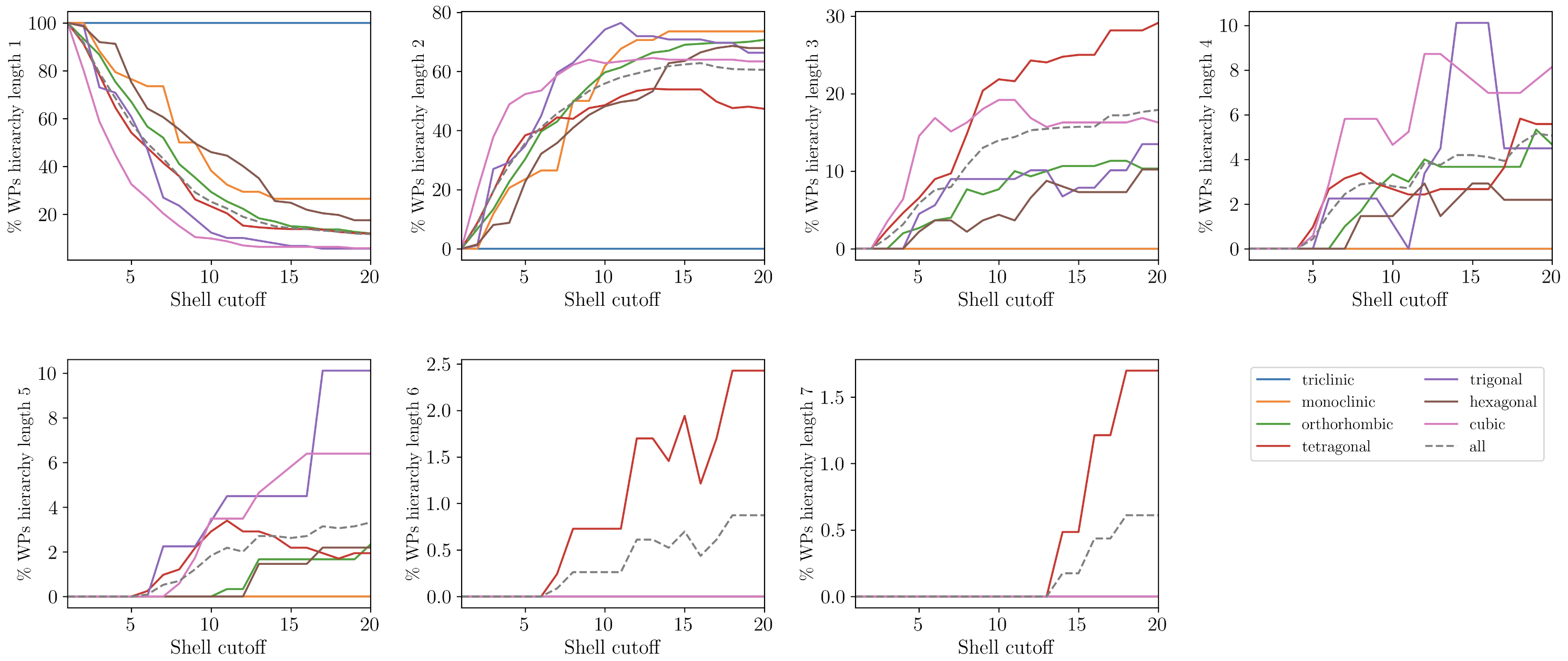}
    \caption{Plots presenting an overview of symmetry enhancement/breakdown across different crystal classes. Each panel shows the percentage of Wyckoff positions from a given crystal class with a hierarchy of length $n$ (from one to seven in reading order) as a function of shell cutoff $-$ referring to the number of space groups and Wyckoff positions that specify the symmetries of bonds for each and every shell $\leq n$. The different solid lines in each plot refer to one of the seven crystal class and the dashed line is the aggregate of the information across all crystal classes.}
    \label{fig:stats}
\end{figure*}

 This information is summarized in Fig.~\ref{fig:stats} which shows the percentage of Wyckoff positions with hierarchies of different lengths as the shell number increases. Here the assessment of symmetry is made on the basis of fixed single shell numbers while the plots reflect the number of distinct such symmetries that appear up to and including shell $n$. The curve are not monotonic as, for example, all hopping models trivially have length one hierarchies when considered at shell number one, while the hierarchy lengths may increase or decrease as another set of symmetries is added - the latter as the length $p$ hierarchy is removed from the length $p$ list when it is promoted to length $p+1$.   The triclinic class only has hierarchies of length $1$. For all other crystal classes, the percentage of length $1$ hierarchies falls off rapidly with shell number indicating the high frequency of this effect. With the orthorhombic class we start to see hierarchies of length three, four and five (green). The tetragonal class (in red) is the most spectacular with some hierarchies of length seven appearing at high enough shell number. The trigonal, hexagonal and cubic classes all have maximal hierarchy lengths of five and it is the cubic class for which there is the most precipitous fall-off of length one hierarchies. 
 
\begin{figure*}[h!]
    \centering
    \includegraphics[width=0.8\columnwidth]{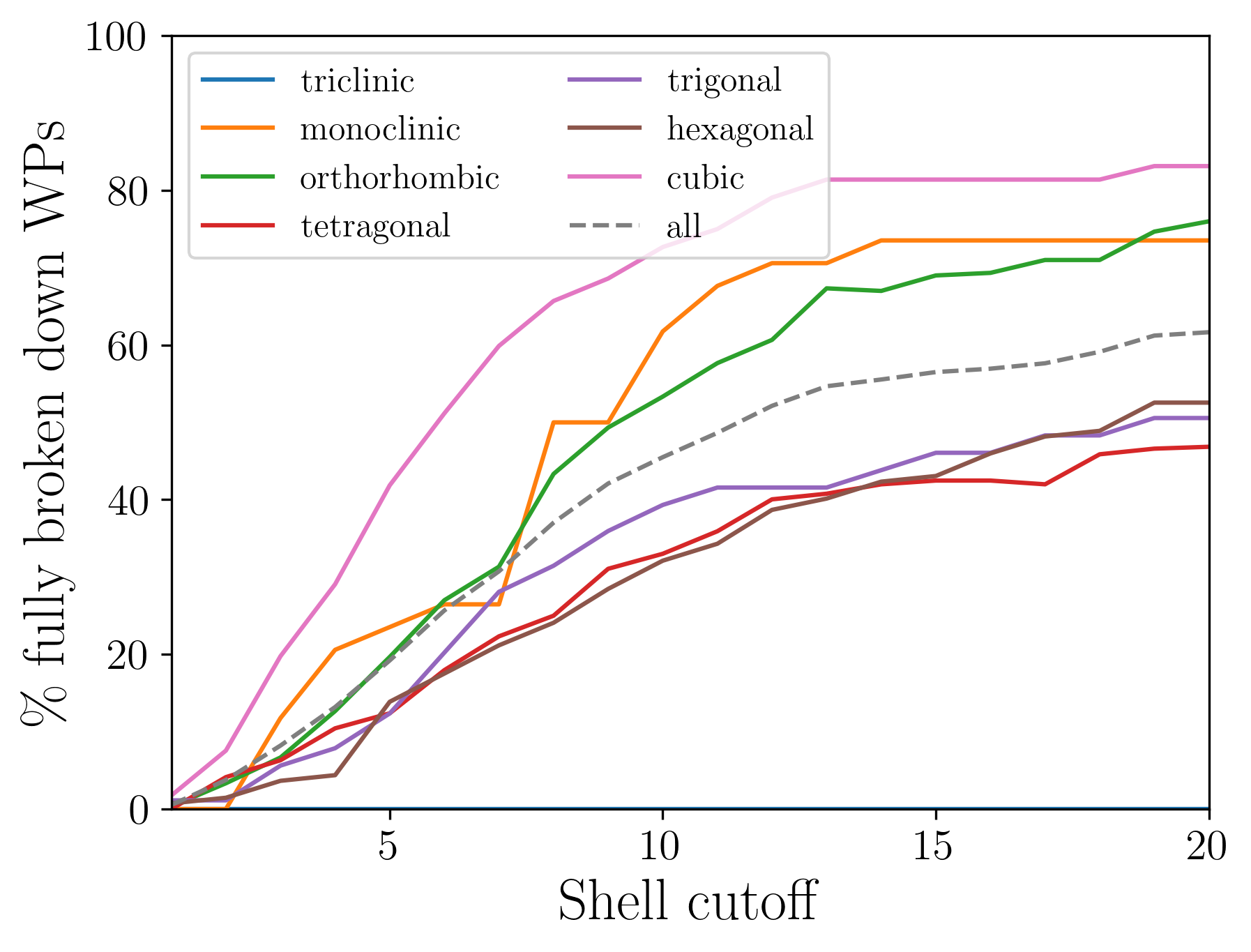}
     \includegraphics[width=0.8\columnwidth]{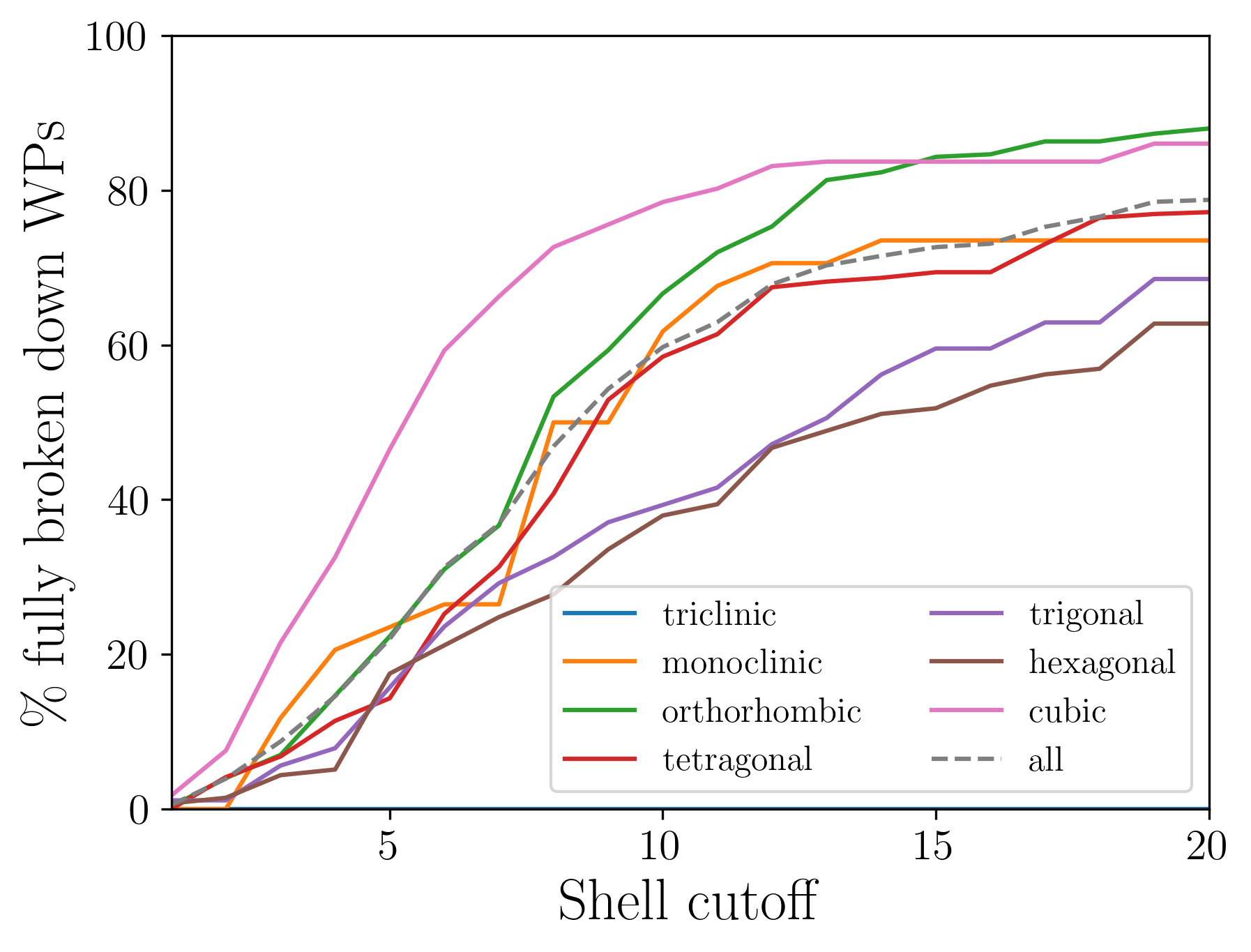}
    \caption{Percentage of fully broken down Wyckoff positions by a certain shell cutoff. The left panel is for shell-by-shell breakdown, the right panel is for shell-cumulative breakdown.}
    \label{fig:stats2}
\end{figure*}

Lengths of hierarchies tell us nothing about enhancement at long-range hopping as the SG \# 1 example illustrates. The final column of Table~\ref{tab:overview} gives statistics that are perhaps surprising. A high percentage of Wyckoff positions have hopping models that do not break down to the parent symmetry even out to the $20$th shell. The ten percent for the triclinic case merely reflects the single Wyckoff position for which there is enhancement at all shell numbers but, for example, the $36.3\%$ for the tetragonal class amounts to $204$ Wyckoff positions! Symmetry enhancement of $s$-wave tight-binding models is far from rare and it can persist even at long-range. This result prompts the question of whether the enhancement persists to infinite range hopping. The plots in Fig.~\ref{fig:stats} already are at least suggestive that there is some convergence in the {\it total set} of symmetries possible for a given set of parent data up to shell number $20$. Fig.~\ref{fig:stats2}(left panel) records the percentage of Wyckoff positions that have a shell that breaks down to the parent data out to the $n$th shell. This too reveals some degree of convergence. But to establish whether an enhanced symmetry persists to longer range we have carried out a direct search to long range hopping for a subset of the exceptional cases up to shell $20$ finding persistence of the enhanced symmetry. Then, in Appendix~\ref{sec:allshells} we have a proof for a single instance of symmetry enhancement to the eigensymmetry for all shell numbers. This is a proof of principle that the phenomenon explored in this paper can indeed survive to arbitrary range hopping length. 
 
\begin{figure*}[h!]
    \centering
    \includegraphics[width=2\columnwidth]{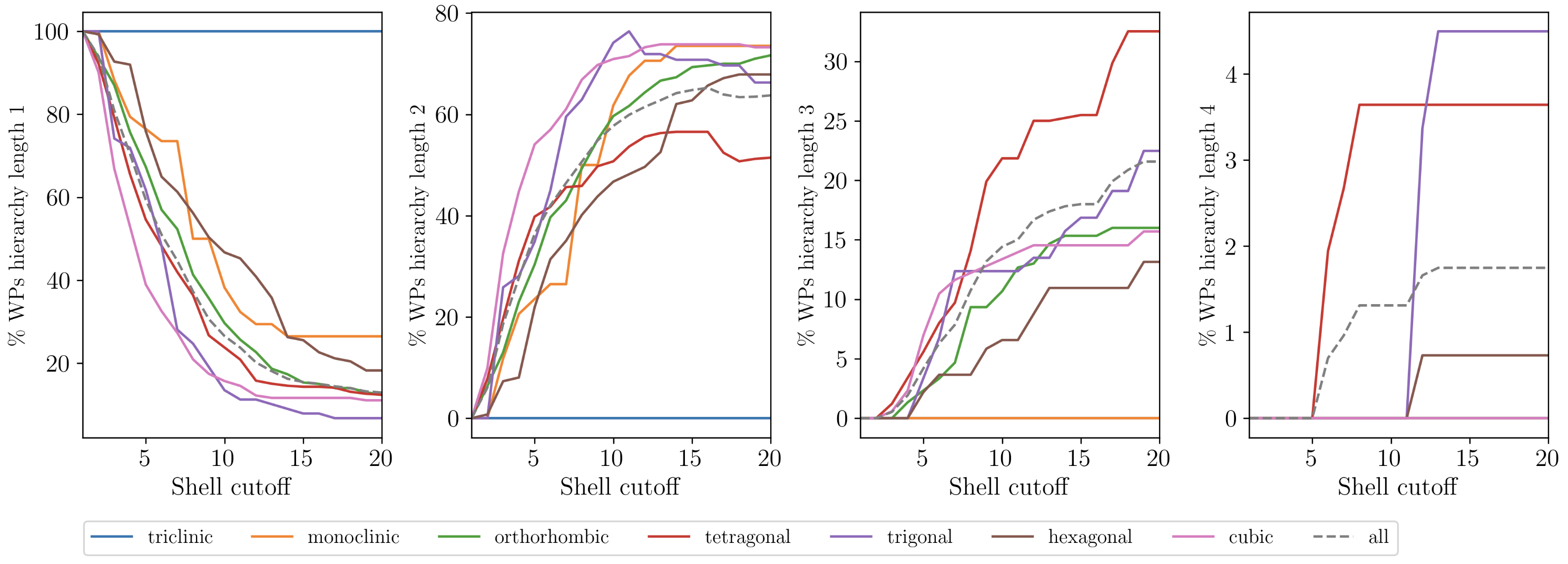}
    \caption{Plots presenting an overview of cumulative symmetry enhancement/breakdown across different crystal classes. By ``cumulative" we mean that, in contrast to Fig.~\ref{fig:stats}, this figure is calculated from the symmetries of $s$-wave tight-binding models including hopping on all shells $\leq n$. Each panel contains the percentage of Wyckoff positions with hierarchy of the specified length as the shell cutoff $n$ is varied and resolved by crystal class. The different solid lines in each plot refer to one of the seven crystal class and the dashed line is the aggregate of the information across all crystal classes.}
    \label{fig:cumulativestats}
\end{figure*}

Exploring symmetry enhancement at fixed shell number is conceptually convenient but from the perspective of physical tight-binding models it is of more interest to consider the cumulative effect of including hopping on shells up to and including shell number $n$. We have examined this problem and statistics are presented in Fig.~\ref{fig:cumulativestats}. We would naturally expect that combining shell numbers would reduce the richness of possible symmetry enhancement $-$ both the proportion of hopping models that remain symmetry enhanced out to the $n$th shell and the length of hierarchies. But the effect of symmetry enhancement remains. Indeed, the simple sample of SG $\#1(1a)$ being enhanced to a single symmetry at all shell numbers tells us that this must also be the case in the cumulative problem. But, Fig.~\ref{fig:cumulativestats} reveals that, indeed, the maximum hierarchy length is reduced from seven in the shell-by-shell case to four in the cumulative case with tetragonal, trigonal and hexagonal all exhibiting examples of length four hierarchies. Then Fig.~\ref{fig:stats2}(right panel) shows that the percentage of cases where symmetries are fully broken down to the parent symmetries is generally higher, shell-by-shell, than for the shell-by-shell case there is not a dramatic difference between them. And, in particular, there are still many cases even in the cumulative case where the symmetry is not broken down at shell $20$.

We also make some remarks about the kinds of symmetry breaking, from the eigensymmetry, observed for different hopping models.  We have observed that there can be breaking of point group symmetries (such as loss of a rotation symmetry) as well as translation symmetry breaking and, indeed, both together. A proxy for translational symmetry breaking is the Wyckoff multiplicity of the enhanced group relative to the parent. Table~\ref{tab:overview} gives a guide to the percentage of multiplicity one Wyckoff positions. These already have high translational symmetry that cannot be broken down further. But these percentages are small for most crystal classes suggesting that translational symmetry enhancement may be quite common. 

Table~\ref{tab:multiplicity} gives a list of the Wyckoff multiplicities that can be reached from a given parent multiplicity. As the symmetry can only stay the same or increase starting from the parent, the Wyckoff multiplicity cannot exceed that of the parent. The multiplicity may decrease however as different Wyckoff positions may be identified under the enhanced symmetry. Symmetry enhancement generally does not allow multiplicities to explore all of the {\it a priori} available values. Of the $17$ possible multiplicities, some do not change. These are $1$ (trivially), $9$, $36$ and $192$. Multiplicities of $24$ or higher may changes only by a factor of $8$ presumably coming from a factor two reduction in the cell size in all three dimensions to preserve the crystal class. 

As we have discussed, the connection between point group symmetry breaking and the number of bond equivalence classes is controlled by a group-subgroup index. This is merely frequently the case for cases where translational symmetry changes. In Appendix~\ref{sec:exceptions} gives some explicit examples of translational symmetry enhancement by changing the hopping range that highlights some of the subtleties involved in gauging how the number of bond colors will change when the size of the primitive cell changes. In this context, it may be interesting to investigate further the detailed origins of the multiplicity translations recorded in Table~\ref{tab:multiplicity}.

\begin{table}[h!]
\centering
\begin{tabularx}{0.9\columnwidth}{| >{\centering\arraybackslash}X | >{\centering\arraybackslash}X |}
\hline
\vspace{1pt}
Wyckoff multiplicity of parent & \vspace{1pt} Wyckoff multiplicity of hopping model \\
\hline
          1 &    1 \\
          2 &    1,2 \\
          3 &    1,3 \\
          4 &    1,2,4 \\
          6 &    2,3,6 \\
          8 &    1,2,4,8 \\
          9 &    9 \\
          12 &    6,12\\
          16 &    2,4,16 \\
          18 &    9,18 \\
          24 &    3,24 \\
          32 &    4,32 \\
          36 &    36 \\
          48 &    6,48 \\
          64 &    8,64 \\
          96 &    12,96\\
          192 &    192 \\
\hline
\end{tabularx}
\label{tab:multiplicity}
\caption{Table giving the possible multiplicities of the hopping model starting from the multiplicity of the parent. }
\end{table}

\subsection{Beyond lattice symmetries}
\label{ssec:beyond}

Our scheme for investigating tight-binding models rests on finding their lattice symmetries. We observed that these are bounded from below by the parent space group and from above by the eigensymmetry group and are connected by a network of group-subgroup relations. But the tight-binding models have more features than can be captured by the lattice symmetries alone. In this section, we describe some of these features and comment on their consequences for the band structures computed from the hopping models. There are two main points to address: (i) the dimensionality of the hopping including whether the hopping percolates across the lattice at all and (ii) whether there is a sublattice selectivity of the hopping. 

For crystal classes of relatively low symmetry, we can expect certain shell numbers to lead to sub-dimensional hopping. Consider, for simplicity, members of the tetragonal crystal class. As these distinguish one axis we can expect different shell numbers to generate one-dimensional hopping  $-$ where the shell connects neighbors along the distinguished axis $-$ and two-dimensional hopping perpendicular to that axis. As the rutile example (Section~\ref{sec:overview}) demonstrates, the connectivity of certain shells is fully three dimensional. A similar observation holds for hexagonal crystals. Triclinic, monoclinic and orthorhombic crystal classes naturally lead to one dimensional hopping models for any given shell. The most symmetric case is also exceptional $-$ cubic crystals tend to lead to fully three-dimensional hopping models. Sub-dimensional hopping directly affects the band structures leading to momentum independent bands in one or two directions in the zone. This, of course, means that the actual symmetry of these models is higher than the lattice symmetry belonging to one of the wallpaper or line groups. In other words, the symmetry data we have provided assumes an embedding in the crystal as is always the case in materials, thus providing information about how the decoupled chains or planes are related to one another.

Having discussed cases where the tight-binding model percolates only in one or two dimensions in the three dimensional crystal, we separate out the zero dimensional case or the possible appearance of finite clusters. Once again this is physics that is not fully captured by the symmetry analysis as the symmetry assignment provides information both about the symmetry of the clusters $-$ which must be one of the crystallographic point groups $-$ as well as how different clusters are embedded into the crystal. A hopping model on a cluster has a finite discrete spectrum where the number of distinct levels is equal to the number of vertices in the cluster. This is the realm of crystals of molecules and it is interesting to see how such structures may come about, not by thinking about how such molecules can be stacked, but instead how restricting hopping to shells can lead to such molecules.  

\begin{figure}[h!]
    \centering
    \includegraphics[width=0.7\columnwidth]{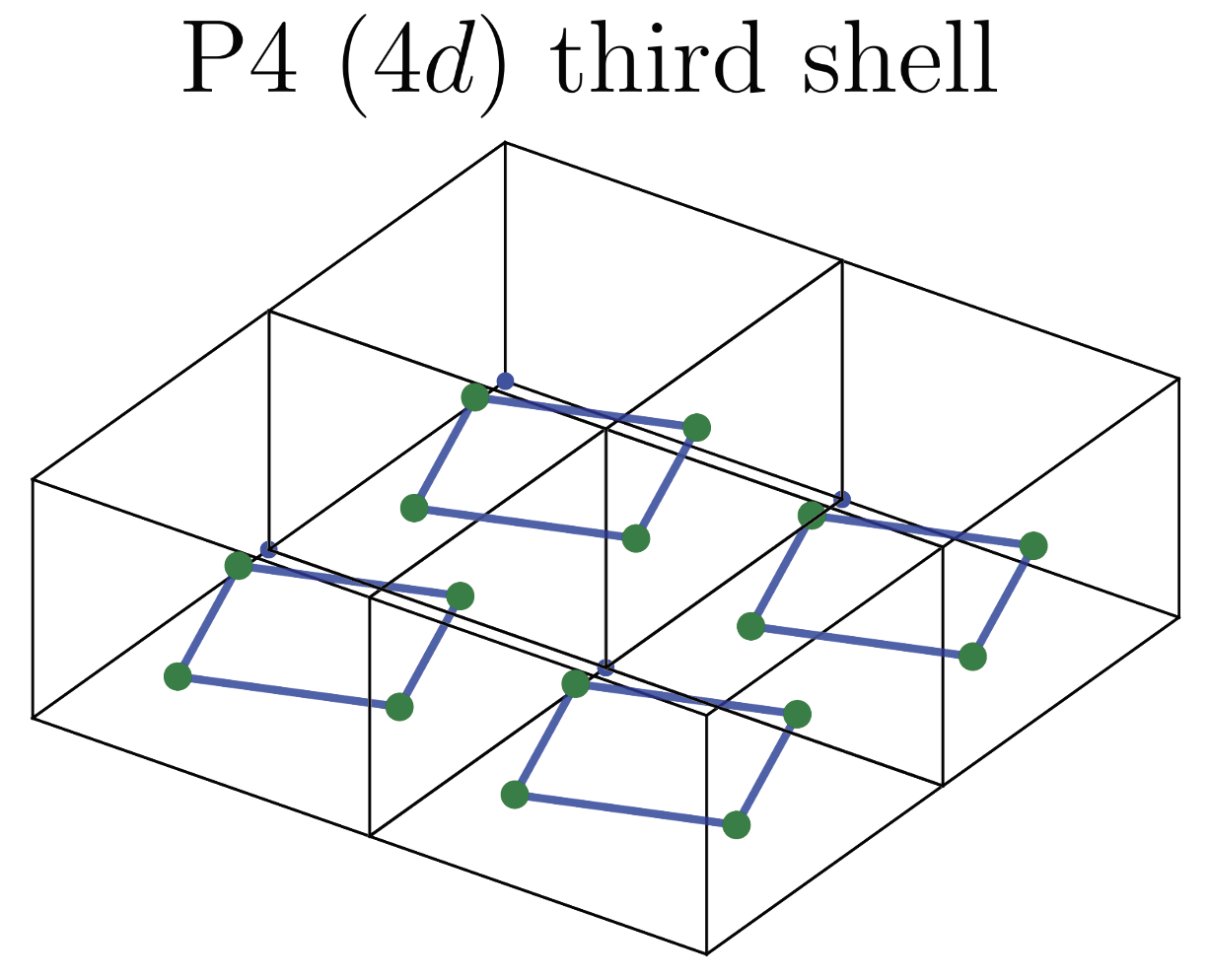}
    \caption{Square clusters of bonds in the third shell of Wyckoff position $4d$ of space group P4.}
    \label{fig:cluster}
\end{figure}

An example of this clustering of bond equivalence classes arises when looking at the third shell of Wyckoff position $4d$ with space group P4. The resulting graph exhibits clusters of 4 bonds which join together to form disjoint squares (Fig.~\ref{fig:cluster}). The tight-binding model resulting from such a hopping model will have an extensive degeneracy due to the nature of the hopping bonds which form clusters similar to molecules. 

Another important class of cases to consider is where, at fixed shell number, the hopping breaks up the Wyckoff set into disjoint sets such that hopping occurs within each set and not between them. This is distinct from the cases already considered as the hopping model within each set may or may not percolate. To be concrete, suppose the Wyckoff multiplicity is two. Then it is typical for a given shell either to connect identical sublattices $A-A$ and $B-B$ or to mix the sublattices $A-B$. The Hamiltonian is therefore either diagonal or off-diagonal in the sublattice basis and not both. This supplies an additional constraint on the model that affects the dispersion relations.

\subsection{Band Structures}
\label{ssec:bands}

\begin{figure*}%
\centering
\vspace{-3ex}
\subfloat[Shell $1$: $123(1c)$]{\includegraphics[height=4.8cm]{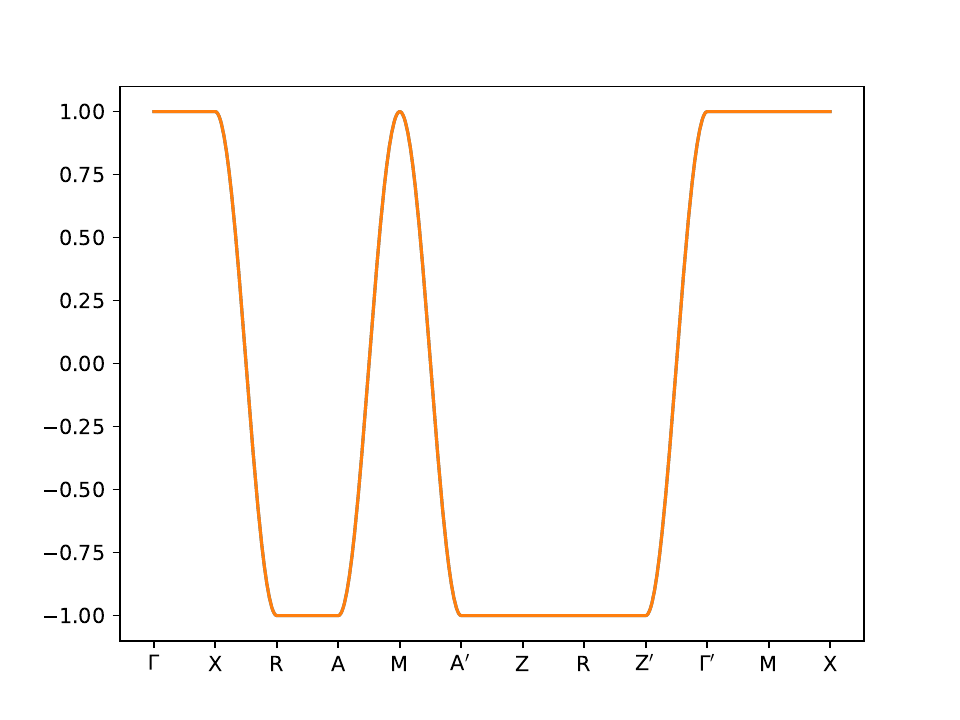}}
\subfloat[Shell $2$: $123(1c)$]{\includegraphics[height=4.8cm]{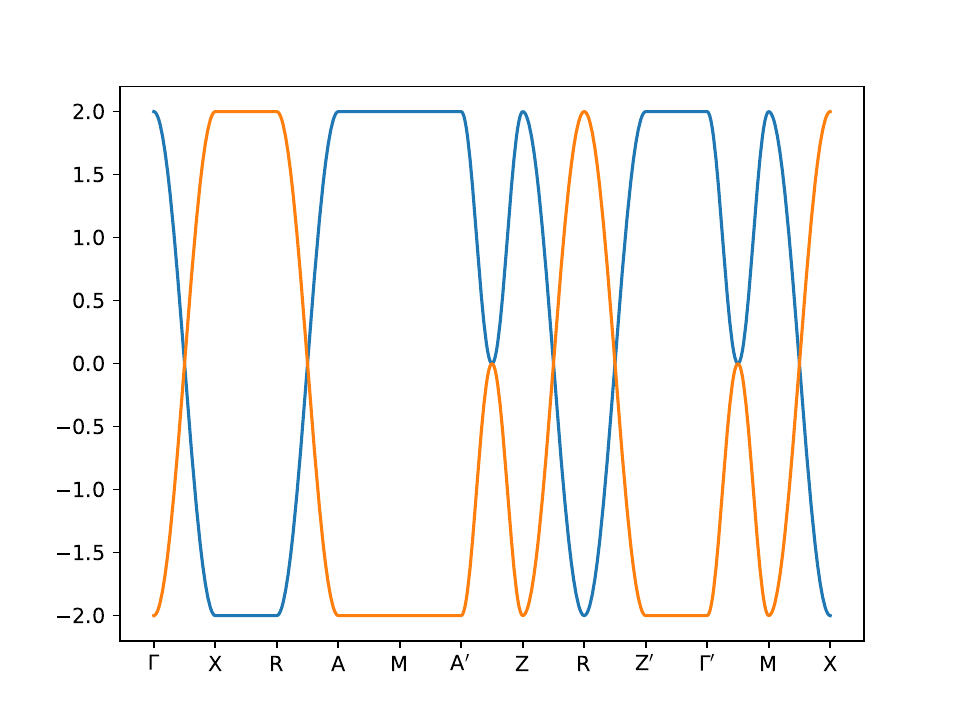}}
\subfloat[Shell $3$: $125(2c)$]{\includegraphics[height=4.8cm]{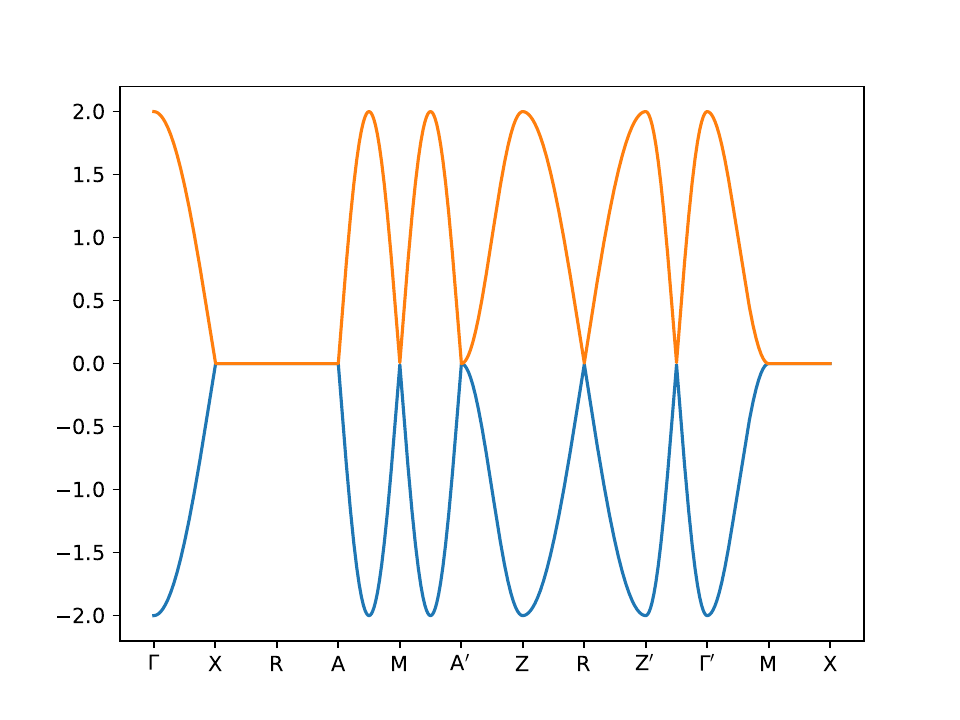}}\\[-3ex]
\subfloat[Shell $4$: $123(2f)$]{\includegraphics[height=4.8cm]{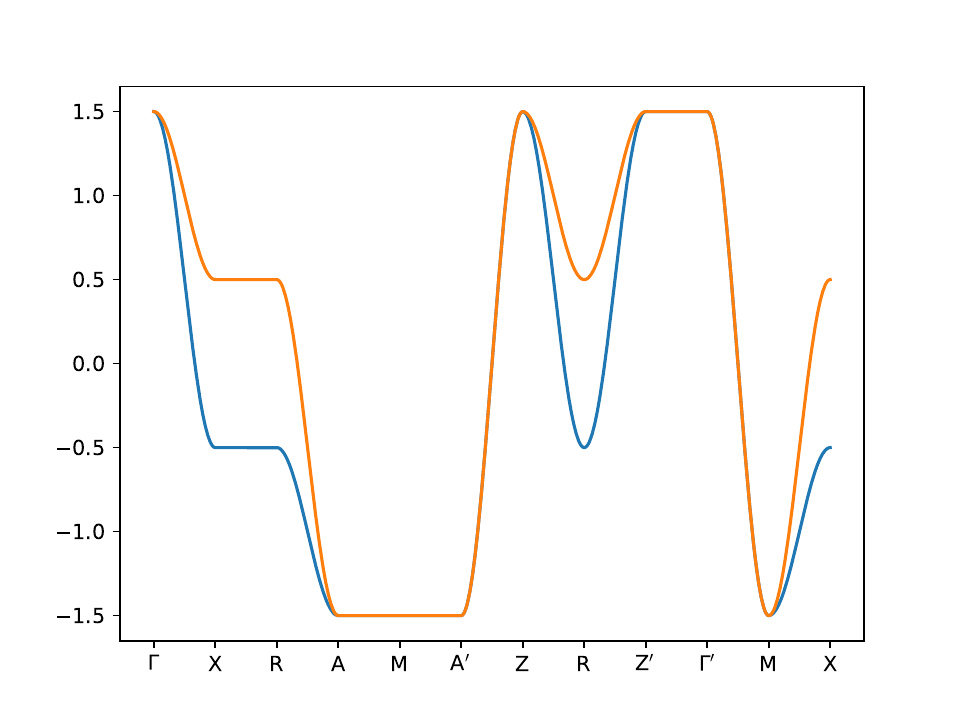}}
\subfloat[Shell $5$: $123(2f)$]{\includegraphics[height=4.8cm]{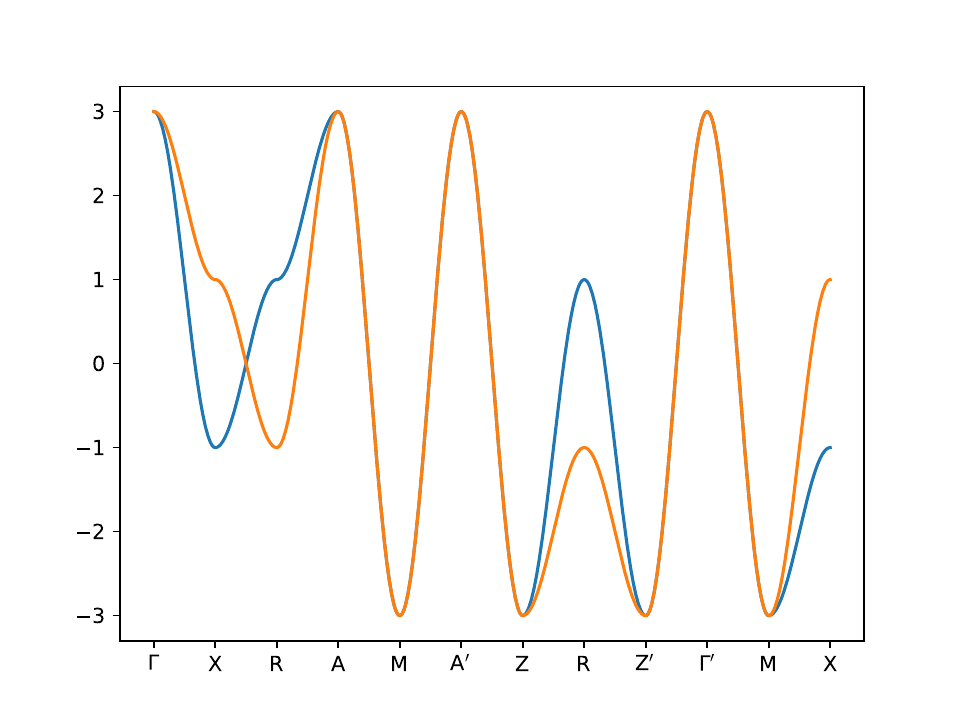}}
\subfloat[Shell $6$: $123(1c)$]{\includegraphics[height=4.8cm]{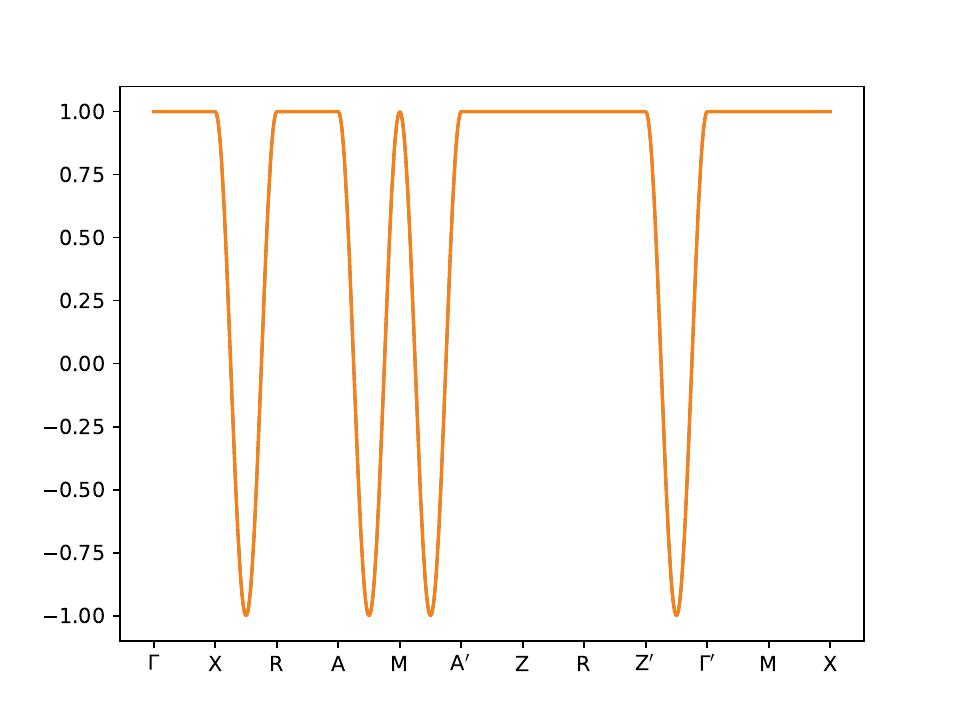}}\\[-3ex]
\subfloat[Shell $7$: $127(2d)$]{\includegraphics[height=4.8cm]{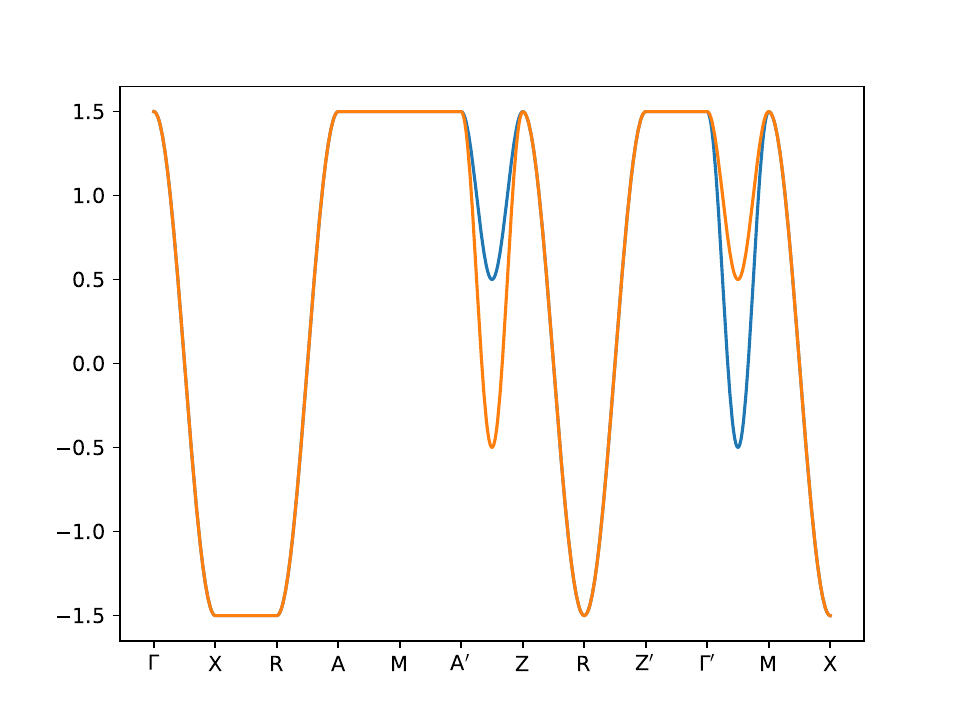}}
\subfloat[Shell $8$: $125(2c)$]{\includegraphics[height=4.8cm]{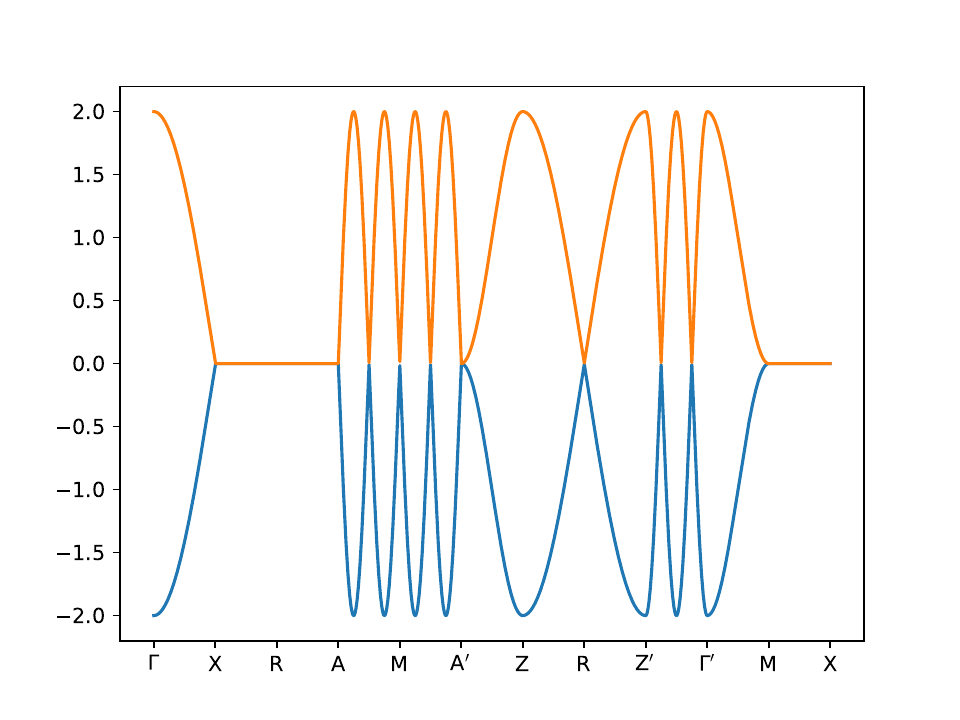}}
\subfloat[Shell $9$: $127(2d)$]{\includegraphics[height=4.8cm]{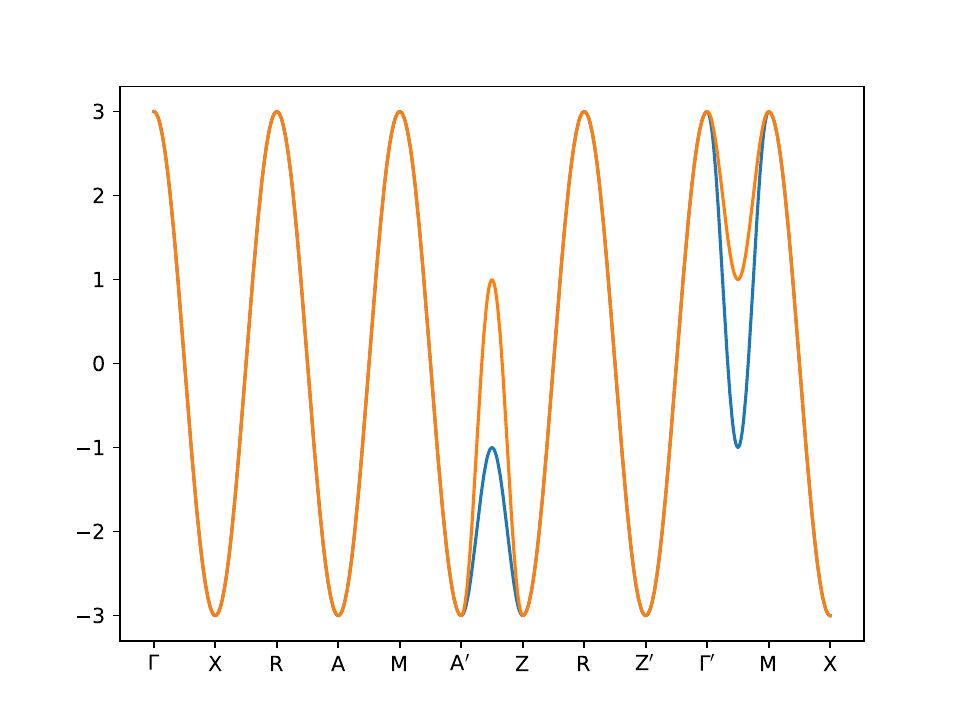}}\\[-3ex]
\subfloat[Shell $10$: $83(1c)$]{\includegraphics[height=4.8cm]{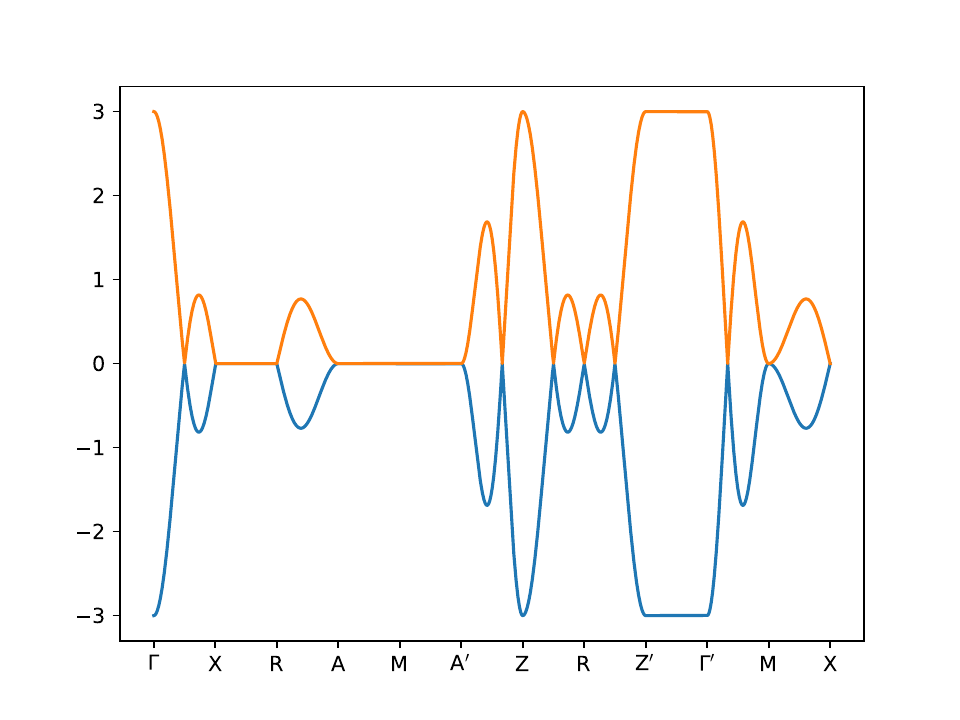}}
\subfloat[Shell $11$: $123(2f)$]{\includegraphics[height=4.8cm]{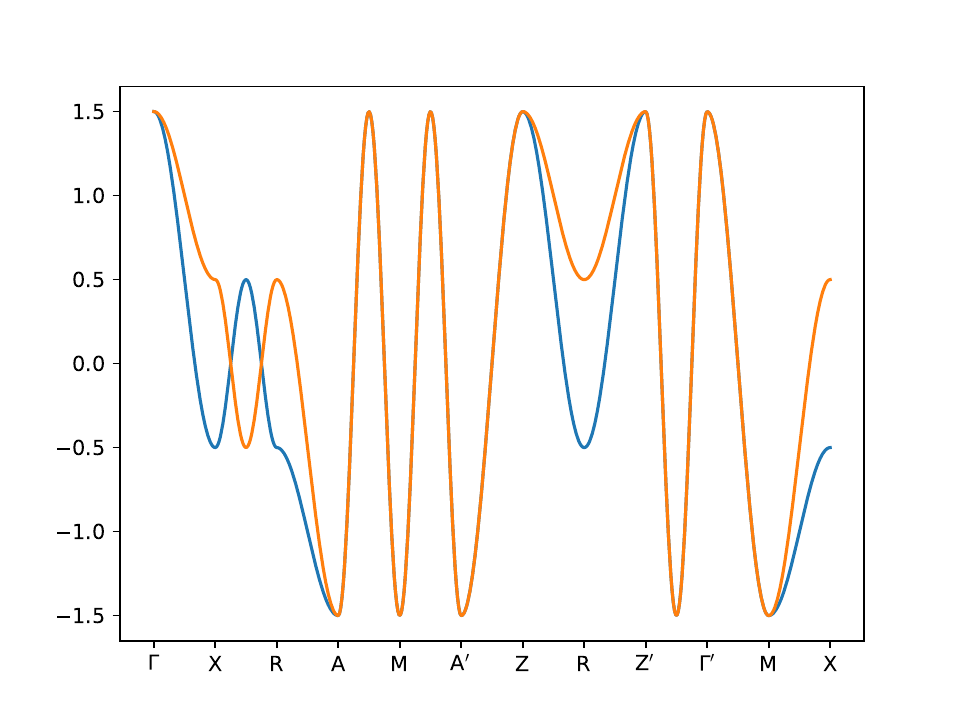}}
\subfloat[Shell $12$: $85(2a)$]{\includegraphics[height=4.8cm]{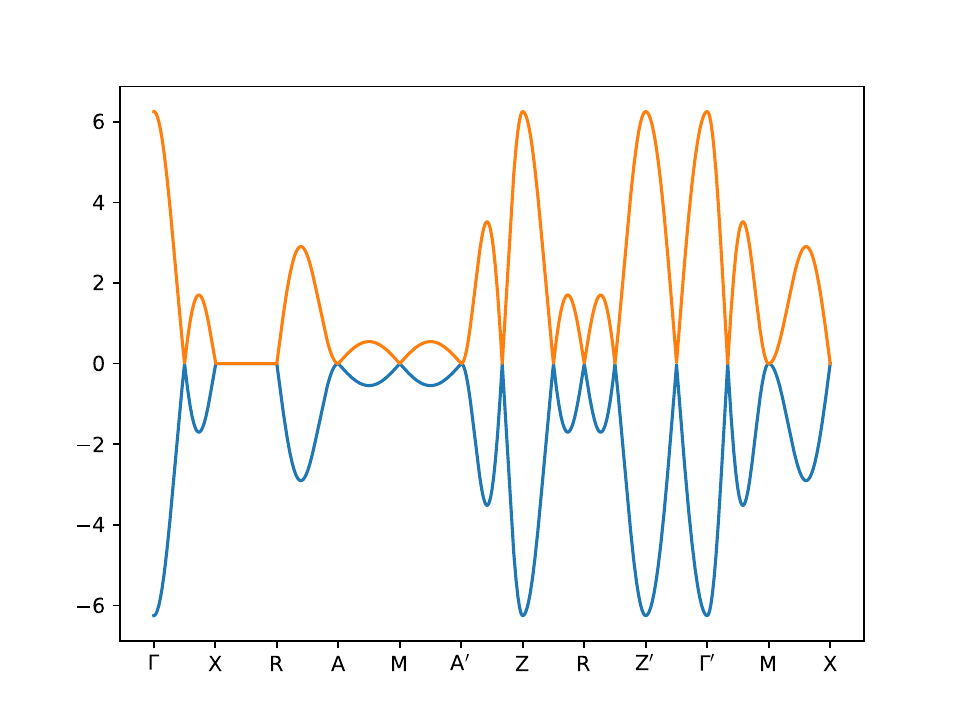}}
\caption{Band structures for tight-binding models at a given shell number for parent space group P4 and Wyckoff position 2c.}
\label{fig:bands}
\end{figure*}

We have computed the band structures for shell number $n$ for $n=1 \ldots 12$ for parent space group P4 and Wyckoff position $2c$. We make this choice because this case exhibits a rich variety of different symmetry groups as the hopping range is changed. The bands along high symmetry directions of the eigensymmetry group $125$ are shown in Fig.~\ref{fig:bands}. We now consider them in turn. 

For shell $1$, the bonds are along $c$ so sublattices $A$ and $B$ decouple and they are identical. The symmetry analysis reveals that the Wyckoff positions merge to a single position. Hence we should expect a single band with emergent symmetries. Indeed the band structure has a single band with a very simple dispersion relation: $\epsilon_{\mathbf{k}}= \cos k_z$. Evidently this has $k_z\rightarrow -k_z$ which one may view as inversion symmetry or two-fold rotation symmetry $2_{100}/2_{010}$ that are not present in the parent group. We note here that the eigensymmetry group is $\# 125$ which is nonsymmorphic. Dropping the partial translations brings us to space group $\# 123$.  As the model at shell one is independent of $k_x$ and $k_y$, the symmetry of the model has all the elements of $\#123$ if we ask which eigensymmetry elements are present. 

Shell $2$ also gives a sub-dimensional hopping model but now with hopping in the ab plane. The band structure in Fig.~\ref{fig:bands}(b) reveals two bands but these actually originate from band folding when the two Wyckoff positions are identified. The model is a simple hopping model on a square lattice with symmetry group $\# 123$. 

Only at shell $3$ do we have a fully three dimensional hopping model. The symmetry group for this model is $125$ demonstrating that sub-dimensional hopping is not a necessary condition for enhanced symmetry. This group has only one-dimensional irreps at $\Gamma$ and $Z$ in agreement with the calculated bands. There are double degeneracies in the band structure along lines $MX$, $RA$ and $XR$ and indeed there are 2D irreps along these directions. These two bands together form a single elementary band representation (EBR). Shell $8$ is identified as having the same symmetry as shell $3$ and the band structures have common features arising from symmetry. 

Shell $4$ has hopping only in the ab plane. The bands are degenerate along $(u,u,0)$. Restricting to inspection of the eigensymmetry elements, one finds that the symmetry group of this tight-binding model viewed as an embedding in three dimensions is $123 (2f)$. While there are 2D irreps at $\Gamma$ and $M$, there are only 1D irreps for this group along $(H,H,0)$, $\Gamma M$ (and also along $AZ$). Shells $5$ and $11$ are identified as having the same symmetry $123 (2f)$ and indeed the band structures have the same qualitative features as for shell $3$. Therefore the puzzle of reconciling the features of the band structure with the computed symmetries is present for all three cases. The puzzle is resolved if we note that this model necessarily has intra-sublattice hopping so the Hamiltonian is diagonal in the sublattice index. But the model has a $2_{110}$ symmetry that swaps the sublattices and that preserves the momentum. This forces the bands to be degenerate along this line. This example, in addition to the subdimensional cases, illustrates the importance of constraints in addition to the basic crystal symmetry constraints. 

Shell $6$, in common with shell $1$, has hopping only along $c$. As in the case of shell $1$ this merges the Wyckoff positions into one and expands the group to $\# 123$. 

Shells $7$ and $9$ are identified as having symmetry $127(2d)$. This correctly identifies the double degeneracies along $XR$ and the splittings along $AZ$ and $\Gamma M$. However, it fails to predict the double degeneracies along $\Gamma X$, $RA$, $ZR$ and $MX$ as there are only 1D irreps along these directions. Once again the solution comes from the intra-sublattice hopping constraint. The group contains a $2_{001}$ symmetry that swaps the sublattices while preserving the momentum along the $(u,0,0)$, $(u,0,1/2)$, $(u,1/2,0)$ and $(u,1/2,1/2)$ directions. This forces the Hamiltonian to be ${\rm diag}(A_{\mathbf{k}}, A_{\mathbf{k}})$. The bands belong to a single EBR even according to the group $127$. 

Shell $10$ has in-plane hopping. It is identified as having different translational symmetry to the parent structure such that the two Wyckoff positions merge. In other words, there is a single independent band. As with shell $2$, the single band appears as two after zone folding. 

This leaves us with shell number $12$. All of the previous shells have either one or two couplings corresponding to the number of bond equivalence classes. To be concrete, shells $3-5, 7-11$ have two couplings. Shell $12$ is exceptional among these in having four couplings that connect sublattices $A$ and $B$. The symmetry is fixed to $\# 85$ and Wyckoff position $2a$. This predicts the appearance of the double degeneracies observed at points  $M$, $R$, $X$, $A$. However, the zero modes along $XR$ are unexpected. Instead there should be two 1D irreps along this line. This time we cannot fall back on intra-sublattice hopping of shells $1,4-7,9,11$ and the constraint of inter-sublattice hopping does not force the off-diagonal hopping to zero along $XR$. 

To deepen our understanding of the conditions for the nodal line along $XR$ we impose symmetry conditions in a minimal way. In particular there is a four-fold symmetry $(k_x,k_y,k_z)\rightarrow (-k_y,k_x,k_z)$ together with complex conjugation of the inter-sublattice hopping term: $A(-k_y,k_x,k_z)=A^*(k_x,k_y,k_z)$ . At most there are four terms corresponding to the four bonds belonging to the same bond equivalence class. Only one of these bonds may be chosen freely: $(\delta_x,\delta_y,\delta_z)$. Now we require that there is a nodal line along $XR$: $(1/2,0,u)$. This produces the condition:
\begin{equation}
e^{-i u \delta_z} \cos(\pi \delta_y) + e^{i u \delta_z} \cos(\pi \delta_x) = 0
\end{equation}
which is satisfied either when $\delta_x = (2p + 1)/2$ and $\delta_x = (2q + 1)/2$ for integer $p,q$ or when $\cos(\pi \delta_x)= \cos(\pi \delta_y)$ and $\delta_z = (2m+1)/2$. The first condition is realized for shell $12$.

\subsection{Application to altermagnetism}
\label{ssec:altermagnetism}

The work in this paper has a direct bearing on altermagnetism especially in insulators. Altermagnets are colinear compensated magnets where the oppositely oriented moments are related by screw or glide symmetries combined with time reversal but not by translation nor inversion symmetries combined with time reversal. This has the consequence that the band structure has an anisotropic spin-splitting in momentum space \cite{Smejkal,Smejkal2020}. The work in this paper directly addresses the following question. What is the shortest range U(1) preserving exchange coupling between pairs of magnetic ions that leads to an altermagnetic splitting? 

Perhaps the simplest example of altermagnetism in three dimensions is derived from the rutile structure described in Section~\ref{sec:overview}. The crystal data for this example is P42/mnm (\# 136) and Wyckoff position 2$a$. The magnetic moments of opposite orientation are placed on the two Wyckoff sites in the primitive cell. These site are related by a $C_4$ rotation in real about the $c$ axis followed by translation through $(1/2,1/2,1/2)$ and the magnetic structure is left invariant by combined these with time reversal. 

Altermagnetism is most cleanly defined in the zero spin-orbit coupled limit such that the moment orientations in spin space play no role in the physics and such that the symmetry is enhanced to include $U(1)$ rotations about the moment direction among other operations that act on spin and real space differently. 

Although altermagnetism in materials would originate from the collective ordering of interacting magnetic moments we may see the essential physics through a tight-binding model of electrons hopping in the background of the magnetic structure. As discussed elsewhere \cite{mcc2024} a tight-binding model respecting the symmetries takes the form:
\beq
H =  - \sum_{n,a} t_{n,a}  \sum_{\langle i, j \rangle_{n,a} } c^\dagger_{i\sigma}c_{j\sigma} + {\rm h.c.} - J \sum_{i} c_{i\alpha}^\dagger \left( \mathbf{S}_i \cdot \boldsymbol{\sigma}^{\alpha\beta} \right) c_{i\beta}
\eeq
where $n$ labels the shells on the lattice and $a$ the inequivalent classes of bonds at a given shell number. The moments $\mathbf{S}_i$ are classical vector spins with fixed anti-colinear orientation.

\begin{figure}[h!]
    \centering
    \includegraphics[width=0.9\columnwidth]{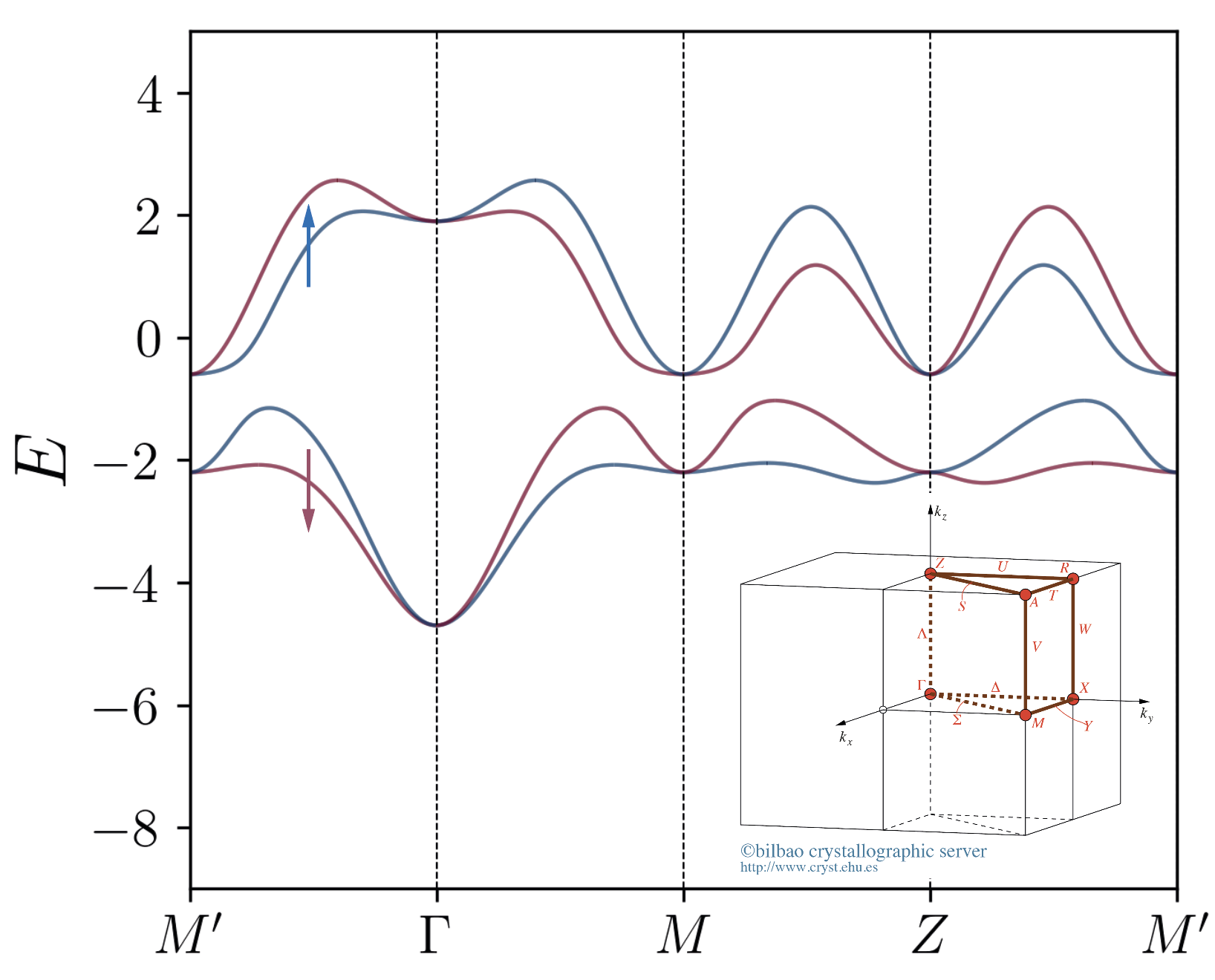}
    \caption{Band structure for minimal tight-binding model on the rutile lattice. The Wyckoff position has multiplicity two and the model has a double exchange term coupling the electrons to moments that are located on the lattice sites and anti-aligned between the two Wyckoff sites.}
    \label{fig:rutilealtermagnetism}
\end{figure}

An illustration of the band structure for sufficiently long-range hopping is shown in Fig.~\ref{fig:rutilealtermagnetism}. Owing to the lack of spin-orbit coupling, the spin projection along the classical moment direction is a good quantum number and labels the bands. The up-spin band is elongated along $[110]$ and the down-spin band by symmetry is elongated along $[1\bar{1}0]$ with the result that there is a $d$-wave spin-splitting in momentum space. This effect boils down to the magnetic and crystal symmetries.  

Now we connect altermagnetism to our study of symmetry enhancement in tight-binding models. To do this, we note that a minimal tight-binding model that produces the spin-split bands of Fig.~\ref{fig:rutilealtermagnetism} has nearest neighbor hopping and hopping between identical magnetic sublattices across the primitive cell diagonals $[110]$ and $[1\bar{1}0]$. The nearest neighbor hopping serves to fully connect the sites and the diagonal hopping induces the altermagnetism. The reason for this is that there are two inequivalent bonds for this shell related by the $C_{4z}$ translation mentioned above. These inequivalent bonds are precisely what allow for the elongation of the bands of different spin. 

Of course, this means that with the nearest neighbor hopping alone there is no altermagnetism even though the lattice symmetries indicate that it should be present. This example generalizes to other lattices admitting altermagnetism. Indeed there are examples such as MnTe $-$ a triangular stacked system $-$ where the inequivalent bonds of fixed shell number arise only as rather long-ranged couplings \cite{mcc2024}. This has physical implications at least for magnetic insulators where altermagnetism is in principle visible as a chiral splitting in the magnon bands \cite{chiralmagnon2023}. In such cases, we might expect the effective hopping or effective exchange couplings to be short-ranged such that any altermagnetism in such systems is likely to be weak. 

This physics is likely borne out in the well-known magnet MnF$_2$ which has the rutile structure. This material that is used in neutron scattering training courses appears to be a classic case of an XXZ antiferromagnet with doubly degenerate magnon bands (albeit with a spectral gap). Inelastic scattering, performed fairly routinely in this magnet, has not uncovered signs of magnon splitting that is admitted by lattice symmetries. 

\section{Summary and Open Questions}

Physicists are familiar with situations where symmetries emerge as an effective description of some phenomenon that are not present at a fundamental level. In this article, we have observed that the symmetry of $s$-wave tight-binding models frequently depends on the range of the hopping integral. One can imagine these insights playing a direct role in effective tight-binding models used to describe materials within some window of energies around the chemical potential. 

Starting from some parent space group and Wyckoff position one may implement all the symmetries of the parent in the tight-binding model at some shell number and find that the symmetries of the resulting model are higher than those of the parent. The symmetries are bounded by those of the eigensymmetry group: the symmetries of the lattice points. We have found that, for fixed parent symmetries, the hierarchy of symmetries as the shell number is changed can include up to seven different cases. These may include both point group symmetry breaking or translational symmetry breaking or both. The principal constraint on the set of symmetries for a given parent is that they belong to the same lattice complex. 

We have explored this phenomenon in full generality for all $230$ space groups and Wyckoff positions and for shell numbers at least up to $20$. For many cases, the symmetry of the tight-binding model does eventually break down to the parent symmetries beyond some shell number. But for a significant fraction of cases this appears not to happen. 

The key concept underlying the breakdown of symmetries from the eigensymmetry is the set of bond equivalence classes. For all the bonds at a given shell number one may implement the parent symmetries and label all bonds covered by these operations by a single color. This defines a bond equivalence class and the presence of multiple colors for a given shell number signals the breakdown of symmetry. This idea can be made precise for cases where the number of colors belonging to the eigensymmetry group equals one and where the translational symmetry is preserved. These conditions are not particularly stringent as they include around $80\%$ of all the cases we explored. We have showed that, when these conditions are met, the number of colors is equal to the group-subgroup index connecting the eigensymmetry and the symmetry of the tight-binding model. We have argued that one should not expect to be able to compute the number of colors from an index in full generality and we have an extensive discussion of how translation symmetry breaking can affect the number of colors.

We have surveyed the landscape of possible tight-binding models and how their symmetry may change with shell number both for cases where hopping takes place on an isolated shell and where hopping takes place on all shells up to some fixed cutoff. We have noted that symmetry alone does not account for the richness of these models as there are many instances of non-percolating shells as well as subdimensional hopping. 

As band structures are strongly constrained by symmetry, the considerations of this paper have a bearing on the physics of non-interacting quasiparticles. One expects some notion of locality for effective tight-binding models in physical settings describing sets of bands that, in aggregate, are topologically trivial. Precisely how the hopping integrals fall off with distance depends sensitively on details but, in all cases, one expects some cutoff beyond which experimental details will be insensitive to inclusion of further shells. Enhanced symmetries of the type discussed in this article may then arise in a wide variety of situations where $s$-wave hopping arises: for example in photonic crystals, spin waves, especially in insulators, phonon dispersion relations and certain sets of electronic bands. Concretely, the bond coloring breaking down symmetries to the parent symmetries directly ties to the exchange splitting necessary to observe altermagnetic splittings e.g. of magnon. Among the features constrained by symmetry is the band topology. Topological quantum chemistry (TQC) supplies information about topology including any symmetry indicator. Our work shows that the basic crystal information is not necessarily sufficient to specify the symmetry of a tight-binding model. However, TQC can be adapted to finite-range hopping models such as those studied here by first identifying the symmetry of the tight-binding model and then inducing a band representation from the trivial site symmetry group appropriate to these tight-binding models. 

Another application of these results is to symmetry broken order as, at some level, we have spelt out possible bond orders corresponding to charge density waves that may arise through spontaneous symmetry breaking from some higher symmetry structure.  

It is natural to ask whether the findings of this paper generalize further to cases with orbital anisotropy and spin-orbit coupling. We have restricted our study to hopping models including single Wyckoff positions and it is an open question to establish the degree of symmetry enhancement when there is hopping between multiple Wyckoff positions. Our study has covered all space groups in three dimensions but one may address similar questions for lower dimensional symmetry groups as well as for magnetic groups. 

One set of open questions raised by our study relates to long-range hopping models. Empirically we have found many cases where the symmetry is not broken down to the parent symmetries up to some large cutoff. We have also demonstrated, for one such instance, that the symmetry is enhanced at arbitrary shell number. It remains to construct a proof for the remaining cases that the symmetry is enhanced for arbitrarily long-ranged hopping. 

Among all the parent groups, Wyckoff positions and shell numbers considered, we have delineated when a generalized index formula computes the number of colors and when it does not. For the large class of cases where the formula always works the result relies only on the orbit-stabilizer theorem. For other cases we have an understanding why one should not expect a simple index formula to exist. One open question is to find the minimal ingredients necessary to compute the number of colors in any instance. 

We have considered the setting of tight-binding models based on $s$-wave hopping. We hope that the concrete implementation of our classification programme will also be of use for analogous efforts in related model families with the potential of exhibiting symmetry enhancement. 

Finally, alongside an exploration of the symmetry of anisotropic tight-binding models, it would be interesting to investigate the effect on symmetry of including interaction terms in the Hamiltonian. The analysis in this paper suffices to demonstrate that, for example, Heisenberg models have a rich set of enhanced symmetries. In fact, the work in this paper carries over directly to models with interactions without multiorbital physics. Extension to other cases is, naturally, of interest.

\begin{acknowledgments}
We acknowledge useful discussions with Andrej Mesaros, Jeff Rau, Judit Romhanyi, Hana Schiff and Masafumi Udagawa. This work was in part funded by the Deutsche Forschungsgemeinschaft under grant W\"urzburg-Dresden Cluster of Excellence on Complexity and Topology in Quantum Matter --\textit{ct.qmat} Project-ID 390858490 - EXC 2147, and grant SFB 1143 (project-id 247310070). 
\end{acknowledgments}

\bibliography{references_tb}

\clearpage
\appendix

\section{Classification algorithm}
\label{app:algorithm}

\subsubsection{Cell convention choice}
To numerically obtain a classification of shell anomalies, one is forced to select a convention on the lattice parameters for each crystal class, and the coordinates for Wyckoff positions which have free parameters. This allows us to build a lattice decorated by bonds which can then be ordered according to length, giving us a clear ordering of the shells. The choice of this convention is physically irrelevant, as long as the lattice parameters are not chosen to be fine-tuned. The only difference among the different conventions is the ordering of the shells. 

\subsubsection{Description of algorithm}

There are a total of $1731$ Wyckoff positions in three dimensions, making it intractable to obtain
a full classification of enhanced symmetries by hand. Instead, using the data and tools in the
Bilbao crystallographic server \cite{bcs1,bcs2}, we have developed an algorithm that identifies the enhanced
space group for all Wyckoff positions, for any shell number. Here we provide a general overview of the steps in the algorithm. The Bilbao Crystallographic Server supplies general positions for all the space groups, the Wyckoff position coordinates and the eigensymmetry group for all Wyckoff positions.

As we have seen the eigensymmetry group $E$ of the initial space group $G$ and Wyckoff position plays an important role in characterizing the possible symmetries of hopping models. If $E = G$, that is if the initial Wyckoff position is characteristic and has no enhanced symmetries
beyond the parent space group, then it follows that the bonds themselves cannot have enhanced
symmetries either. In this case, the enhanced space group and Wyckoff position are the same
as the parent crystal data. If  $E\neq G$, then the Wyckoff positions are non-characteristic, implying that there
is a possibility for symmetry enhancement in the bonds too. 

As the next stage, the lattice is generated from parent space group $G$ and initial Wyckoff
position. In particular, from a suitable set of Bravais primitive lattice vectors deduced from the
crystal class to which $G$ belongs an $N \times N \times N$ unit cell (finite) point configuration is generated. From this, the sites connected at a particular shell number are found.

As the next step, we generate the bond equivalence classes. Two bonds are found to be equivalent
if they map to each under some symmetry in $G$ up to some lattice translation. Consequently,
the equivalence classes are just the distinct orbits of the shells under the action of $G$. These are
generated by applying each element in $G$ to each bond, and checking if the image of the bond is
in some equivalence class already (up to a lattice translation) or not. In the former case, it is added
to that equivalence class, while in the latter a new class is formed containing the original bond.
At this point, we reach another dichotomy. If there is only 1 equivalence class, then the bonds
have the same symmetry enhancement as the Wyckoff positions, so the enhanced space group is
the eigensymmetry group. Likewise for the Wyckoff positions. If instead there is more than 1
equivalence class, then symmetries in $E$ which map elements of different classes get broken,
so the enhanced symmetries form a subgroup of $E$, or at the very least the Wyckoff position
multiplicity will change (if only fractional translation symmetries are broken).

Identifying the enhanced symmetries themselves is complicated by the fact that the Bilbao
Crystallographic server only lists the elements of a space group up to Bravais primitive lattice
translations (i.e. basis vectors of the space group’s standard reference). Hence, if the unit cell
of the eigensymmetry group is smaller than the original space group, the list of symmetries in
Bilbao server will not be complete up to lattice translations of the original group.

To remedy this, we do the following. Firstly, we get all the elements of $E$ as listed on Bilbao.
Using the transformation matrix $T$, provided by the NONCHAR tool on BCS, we check if the Bravais
vectors of $E$, given by $TtT^{-1}$ for $t \in {(1 0 0), (0 1 0), (0 0 1)}$ are fractional in the standard
reference of $G$. If so, they are added to the set $E$. Then, we pair-wise compose together the lattice
translations in $E$ which are fractional in $G$’s standard reference. If this composed translation is
also fractional, it is added to $E$. We re-iterate this process until only trivial lattice translations
can be obtained as new translations by composition, indicating that all fractional translations have
been identified. The full list of translations in $E$ up to lattice translations (of the original group $G$) is denoted as $S$. Next, we check under
which elements of $E$ are the bond equivalence classes closed. These form a set $G'$ of enhanced
symmetries. We let $\bar{E} = E - G$ denote the elements of E which the equivalence classes are not
closed under. Similarly $\bar{S}$ are the translations in $S$ which are broken. We need to check if the
elements in $\bar{E}$ can be composed together to generate new symmetries. However, this can be done
by composing just once the elements of $\bar{S}$ with those in $\bar{E}$ . Indeed,
\beq
\forall g,h \in E, \exists t \in S, g' \in E \ {\rm such \ that} \ gh = tg'
\eeq
where $t$ is potentially fractional in $G$’s standard reference. Consequently, composing two or more elements in $E$ will yield an element in the Cartesian product $S \times E$. So it suffices to check all elements in $S \times E$. Since we are interested in the compositions of elements which the equivalence classes are not closed under, we can restrict ourselves to $\bar{S} \times \bar{E}$. This is because if we compose a symmetry that maps equivalence classes to themselves with another symmetry that does not, the
result is another symmetry that the equivalence classes are not closed under.

\section{Exceptions to the color-index relation}
\label{sec:exceptions}

In the main text we showed that, under some conditions, the number of colors at a given shell is identical to the group-subgroup index connecting the eigensymmetry group and the symmetry group of the tight-binding model at that shell. The conditions are (i) that there is exactly one color for the eigensymmetry group and (ii) the translation index equals one. Under these conditions, the index-color relation is exact. When these conditions are not met, one might expect a simple generalization to work:
\begin{quote}
{
\bf Index$^*$
} The number of colors equals the total group-subgroup index multiplied by the number of colors assuming the eigensymmetry group.
\end{quote}
This embellishment frequently computes the number of colors but there are exceptions.

In this appendix we explore how the formula, {\bf Index$^*$} can fail. The simplest class of exceptions comes from cases where the number of colors in the eigensymmetry group does not equal one. We met such a case on the kagome lattice in Section~\ref{sssection:2D} where, at the third shell, there are two inequivalent classes of bonds $-$ those along chains and those connecting opposite sides of hexagons. These two classes are evidently inequivalent by inspection of the lattice and one should not be surprised that they differ in their behavior as the symmetry is broken down. For {\bf Index$^*$} to work, each color at the eigensymmetry group level should split into the same number of colors. In fact, in the example of Section~\ref{sssection:2D}, one class of bonds split into two colors and the other did not split at all. In this light, it would be surprising if a simple index theorem could compute the number of colors when condition (i) is not satisfied. 

We now concentrate on the more subtle case where condition (ii) fails $-$ namely when the translation symmetry changes between the eigensymmetry group and the symmetry group of the tight-binding model. To illustrate the possibilities we take three examples. In the first example, {\bf Index$^*$} does correctly compute the number of colors. The total index is the product of the point group index and the translation index. In this example, the translation symmetry breaking increases the number of colors in proportion to the change in the unit cell volume. In the second example, the index formula prediction differs from the number of colors by a factor of two. In the third example, there is a factor of four between the index and the number of colors. 

\begin{figure}[h!]
    \centering
    \includegraphics[width=0.8\columnwidth]{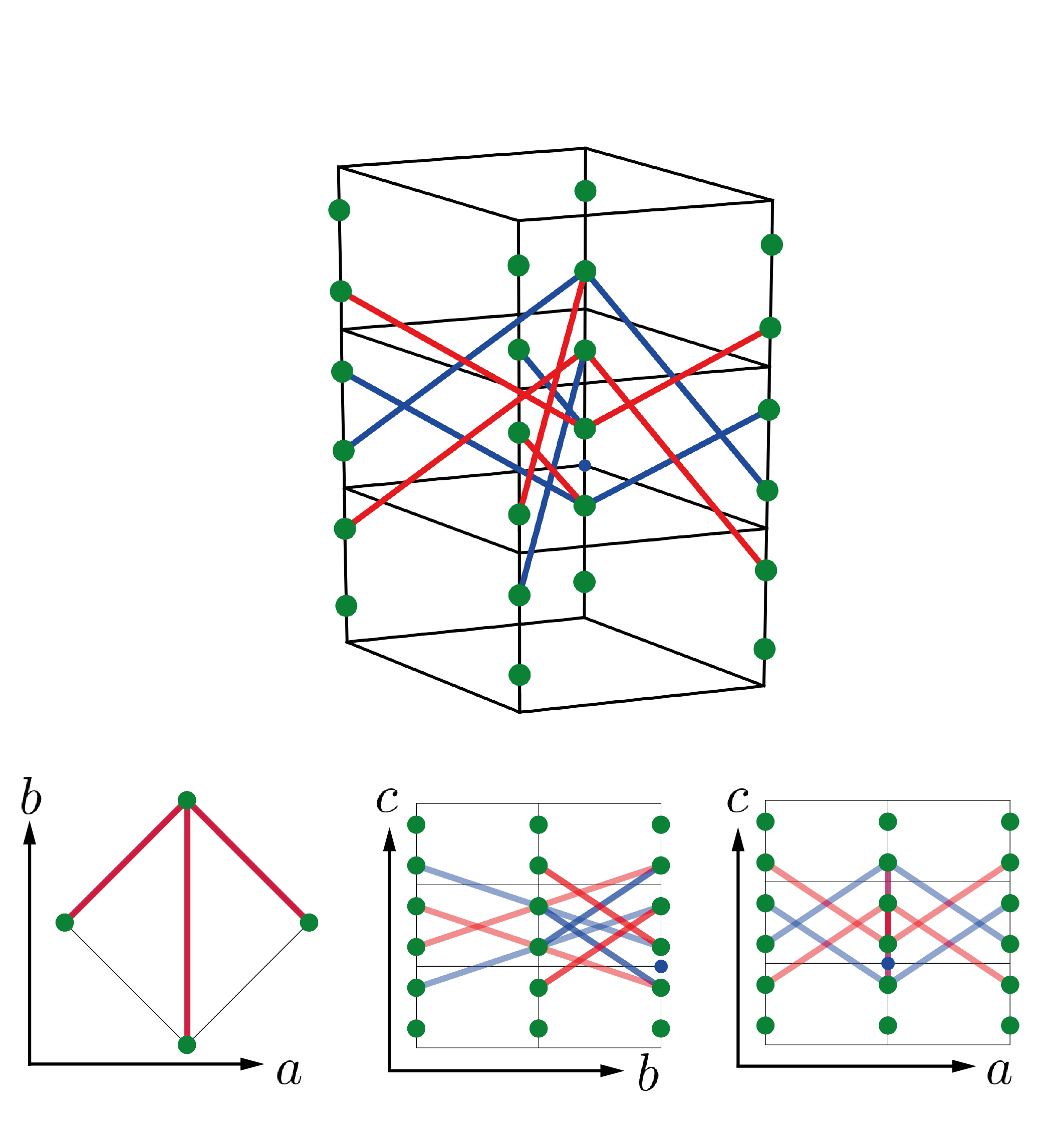}
    \caption{The 2 BECs of $P6_3cm$ at the 5th shell. Here {\bf Index$^*$} correctly calculates the number of BECs}
    \label{fig:P63cm}
\end{figure}

\subsection{Case 1: P6$\mathbf{_3cm}$ at 5th shell}
We begin by considering an example where \textbf{Index*} works. Consider space group $\#185$ and Wyckoff position $2a$ at the 5th shell. The corresponding eigensymmetry group is $\# 191$ with Wyckoff position $1a$, so we see that there is a doubling of the unit cell in the $c$ direction between the two descriptions of the crystal. Naively, one would then expect each BEC at the eigensymmetry group level to split into two BECs due to the quotienting of the primitive translation in along the $c$ direction from the eigensymmetry group. This is exactly what happens and the resulting cell is shown in Fig.~\ref{fig:P63cm}.

\subsection{Case 2: P4/n (4$\mathbf{d}$) at 12th shell}
\begin{figure}[h!]
    \centering
    \includegraphics[width=0.8\columnwidth]{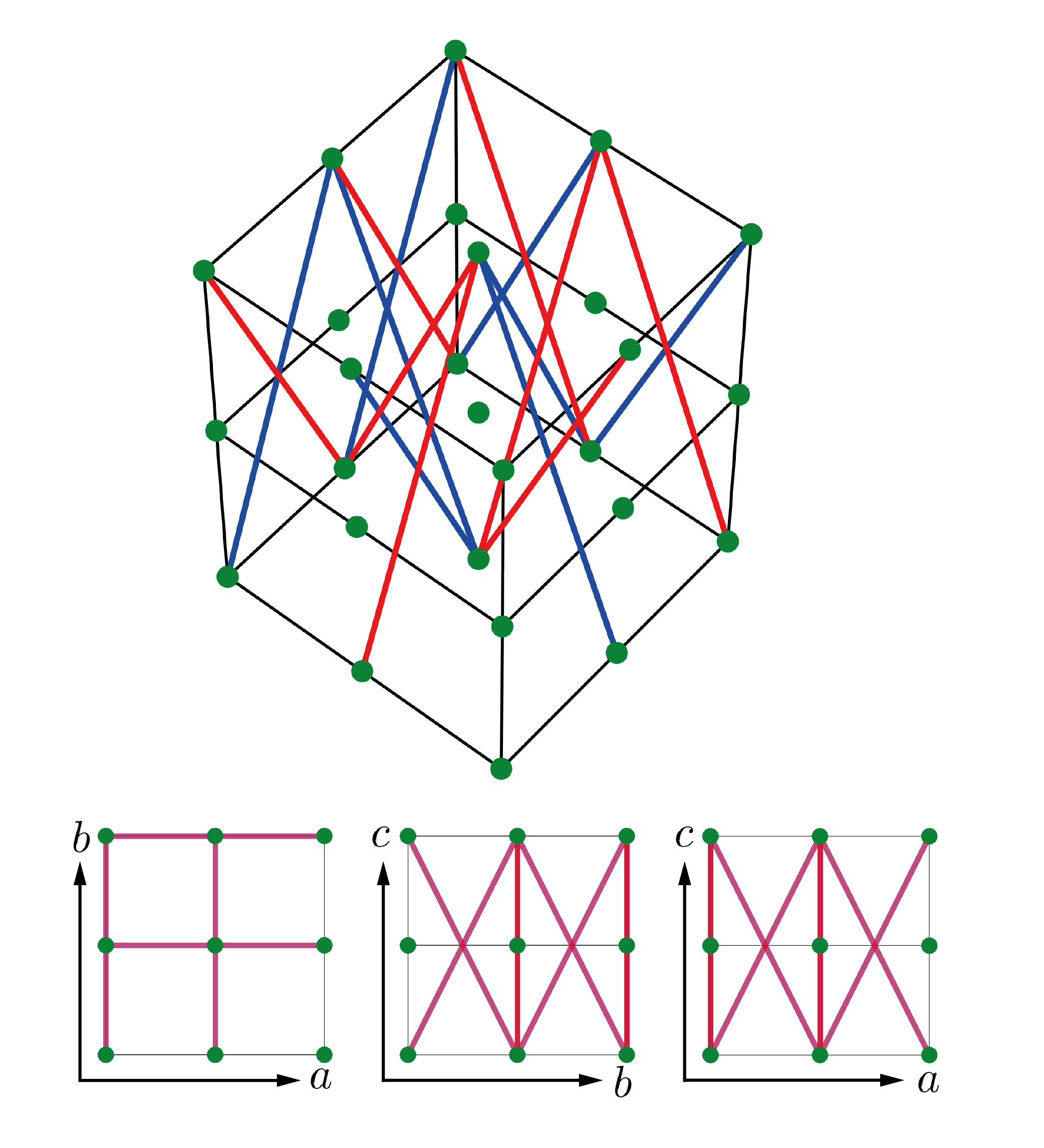}
    \caption{The 2 BECs of $P4/n$ at the 12th shell. {\bf Index$^*$} instead predicts $4$ BECs.}
    \label{fig:P4n}
\end{figure}

Example number two concerns the tetragonal group $\# 85$ and Wyckoff position $4d$ at the 12th shell. A representative bond vector of this shell is $(0.5, 0, 2)$ in units of the primitive unit cell lengths. The lattice complex for the parent crystal data is $\# 123(1c)$, obtained by doubling the unit cell in the $a$ and $b$ directions. Therefore, the group-subgroup index is $i=i_k=4$, coming purely from quotienting out the primitive lattice translations. It follows that the BECs are not symmetric under primitive lattice translations of the eigensymmetry group. Since at the eigensymmetry group level the bonds form a single BEC, we expect each one of these to split into 4 BECs when the translation group is halved, as in the previous example. However, we find that there are actually only 2 distinct BECs. Indeed, for each bond in the shell we note that inversion symmetry (which is present in the parent space group) acts like one of the broken translation symmetries. In other words, every bond effectively experiences an extra translation symmetry coupling it to another bond, and thus the effective group-subgroup translation index for each bond is 2, not 4. As can be seen in Fig. \ref{fig:P4n}, one can show that if the bond vector is of the type $(\pm 0.5, 0 0)$ then the extra translation symmetry is $\{1|\frac{1}{2} \ 0 \ 0\}$, whereas for $(0, \pm 0.5, 0)$ bond vectors the extra translation symmetry is $\{1|0 \ \frac{1}{2} \ 0\}$. This results in only 2 BECs illustrated in Fig.~\ref{fig:P4n}. 

\subsection{Case 3: P4/nbm (4$\mathbf{f}$) at 15th shell}

\begin{figure}[h!]
    \centering
    \includegraphics[width=0.8\columnwidth]{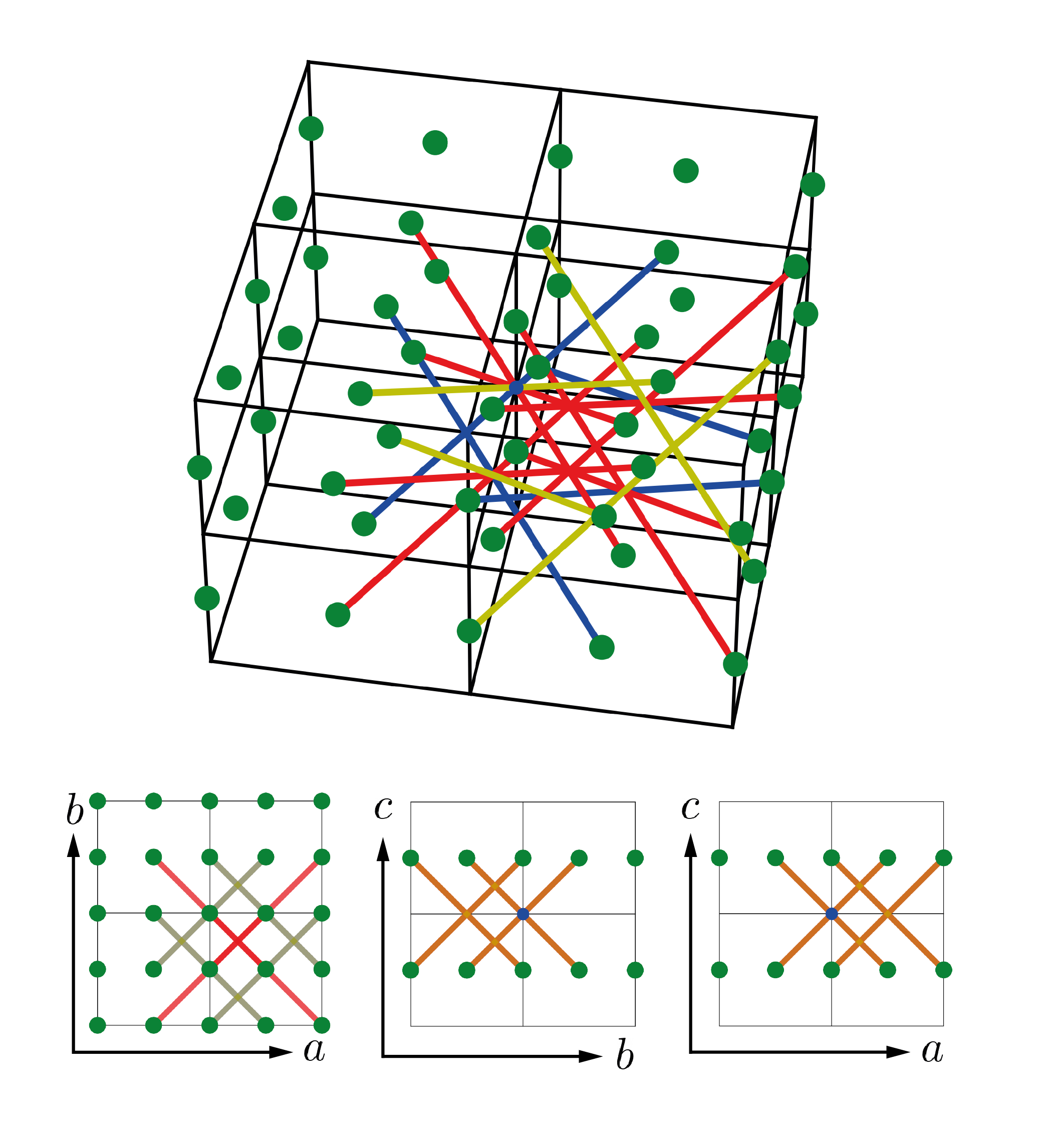}
    \caption{The 3 BECs of $P4/nbm$ at the 15th shell in a case where {\bf Index$^*$} predicts $4$ BECs.}
    \label{fig:125-4f}
\end{figure}

Example number three is for the tetragonal group $\# 125$ and Wyckoff position $4f$ at the $15$th shell. If the primitive cell edge length of the eigensymmetry lattice is one, a representative bond at the $15$th shell is $(1,-1,-1)$. In the eigensymmetry group ($\# 123$) all sites and all such bonds, with coordination number $8$, are identical. With Wyckoff position of multiplicity $4$ and coordination number eight with no bonds shared, we expect $32$ distinct bonds. The number of colors turns out to equal $3$ whereas the translation index equals $4$. {\bf Index$^*$} fails because the index does not equal the number of colors. We address two questions. How does one see the translation symmetry breaking from the point of view of the bonds? And, why are there three colors? To address the first question, we first note that the parent Wyckoff position has multiplicity four while the eigensymmetry group has multiplicity one. So the translational symmetry is broken at the level of the sites. From the point of view of the bonds, Fig.~\ref{fig:125-4f} shows that the pattern of colors emerging from the four sites are different on the different sites reflecting the reduction of symmetry from the eigensymmetry group. Turning now to the number of colors, we note that the parent group has $16$ point group elements (including inversion) so a naive counting would suggest that the $32$ bonds would map into two bond equivalence classes. Closer inspection reveals that the $32$  bonds actually split into $8, 8$ and $16$. In the latter class, each point group element generates a distinct bond. But in the remaining two classes of bonds, inversion and $2_{110}$ play the same role so there are effectively only $8$ elements in the group.  

In general, we notice a pattern in which the BECs at the eigensymmetry group level are each expected to break down into $i_k$ BECs, where $i_k$ is the group-subgroup translation index. However, it is possible that point group symmetries in the parent group can act in the same way as the globally broken translations on the bonds at a given shell. Hence, some of the symmetries may be restored at the level of pairs of bonds. This results in merging of what one would ordinarily expect to be distinct BECs due to the reduction of the translation group.

\section{Example: symmetry enhancement at all shells}
\label{sec:allshells}

We present an example of a crystal lattice whose bonds are always enhanced to the eigensymmetry group. Of course, this implies that the tight-binding model obtained by including bonds of arbitrary range has a higher symmetry than that of the parent crystal. We consider Wyckoff position 1c of space group $P-42m$ (\#111), so the parent crystal data is \#111(1c) corresponding to a simple tetragonal lattice. The associated eigensymmetry group is $P4/mmm$ (\#123) with Wyckoff position 1a. We will now show that the bond-equivalence classes enjoy the full symmetry of the lattice (i.e. the eigensymmetry group), at all shell numbers. 

We begin by listing the elements of $P4/mmm$ and highlight the elements of $P-42m$ in bold. 

\begin{figure}
    \centering
    \centerline{\hspace*{\fill}
    \includegraphics[width=0.6\linewidth]{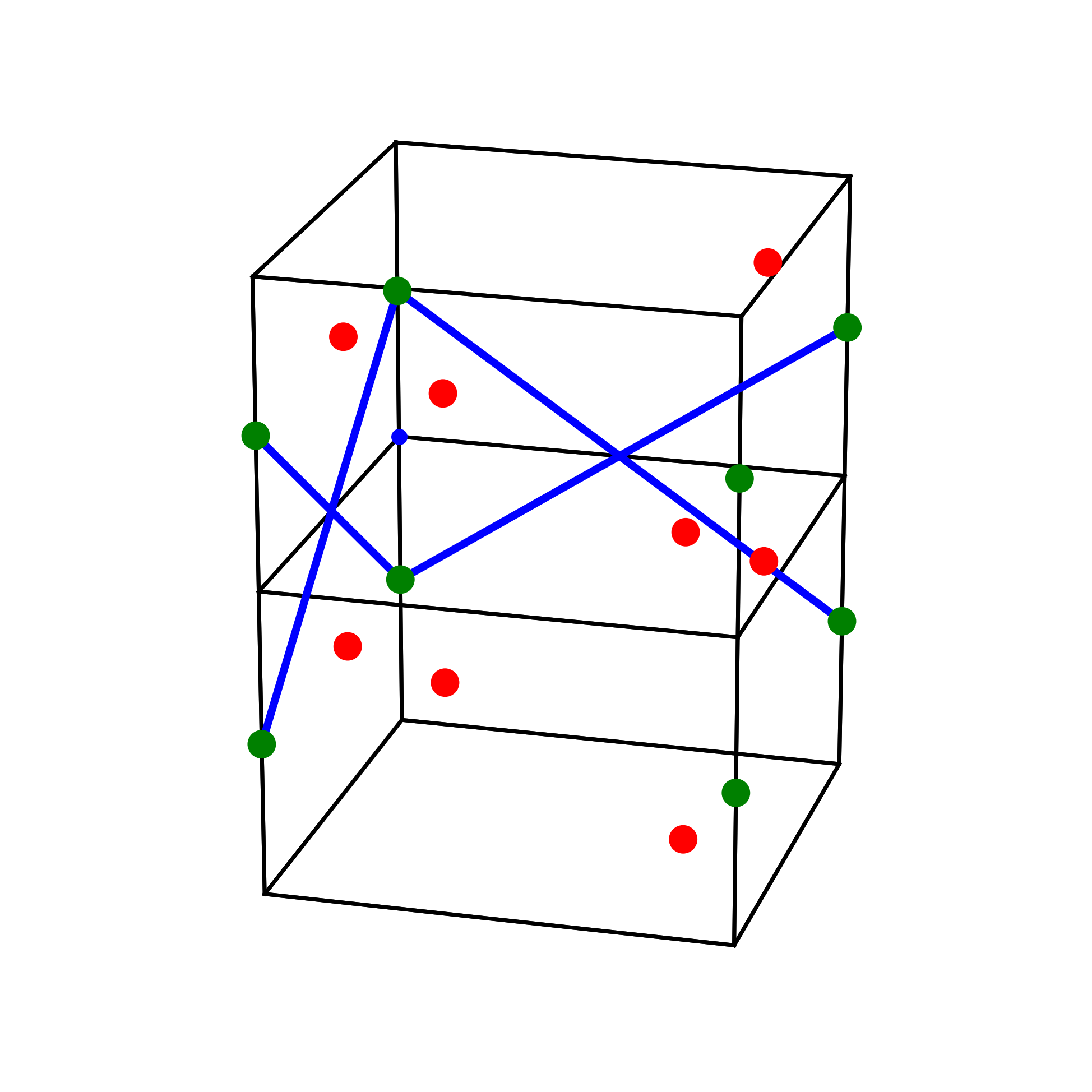}
    \includegraphics[width=0.6\linewidth]{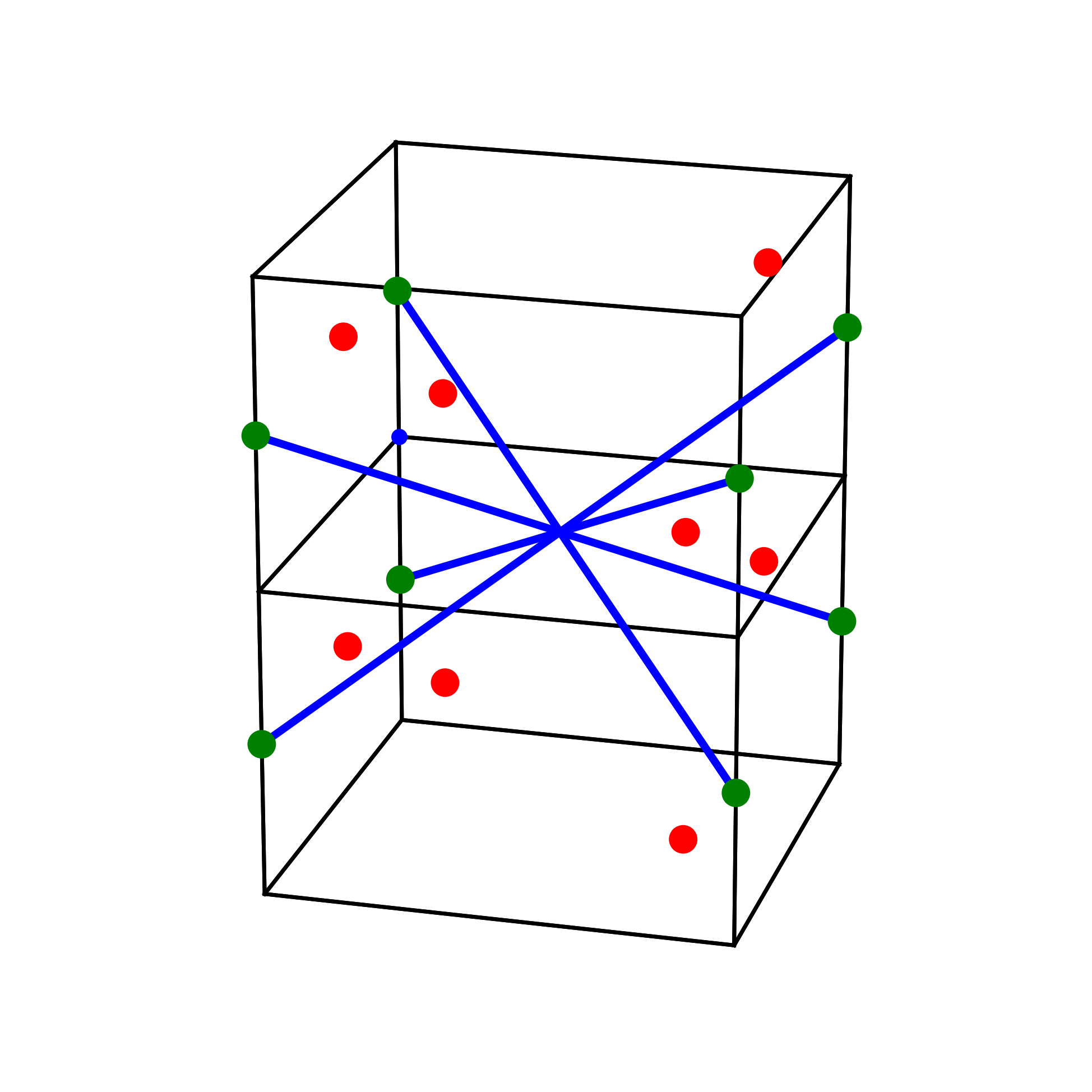}\hspace*{\fill}}
    \caption{Shell number 3 (left) and 6 (right) of a crystal with Wyckoff positions $1c$ (green sites) and $4n$ (red sites) of space group $P4-2nm$, exhibiting a spurious inversion symmetry not present in the parent crystal.}
    \label{fig:111-1c4n}
\end{figure}

\begingroup
\begin{widetext}
\begin{center}
    \setlength{\tabcolsep}{7pt} 
\renewcommand{\arraystretch}{1.5} 
\begin{table}[h]
    \centering
    \begin{tabular}{cccccc}
    $\boldsymbol{\{1|0\}}$ &  $\boldsymbol{\{-4^{\pm}_{001}|0\}}$  & $\boldsymbol{\{2_{001}|0\}}$ & $\boldsymbol{\{2_{010}|0\}}$ &  $\boldsymbol{\{2_{100}|0\}}$ & $\boldsymbol{\{m_{1\pm10}|0\}}$\\
     $\{-1|0\}$ &  $\{4^{\pm}_{001}|0\}$  & $\{m_{001}|0\}$ & $\{m_{010}|0\}$ &  $\{m_{100}|0\}$ & $\{2_{1\pm10}|0\}$
    \end{tabular}
    \caption{List of elements of $P4/mmm$ expressed in the standard reference of $P-42m$, up to trivial lattice translations. The elements of $P$-$42m$ are highlighted in bold.}
    \label{tab:P-42m}
\end{table}
\end{center}
\end{widetext}
\endgroup
We see that $P$-$42m$ is obtained from $P4/mmm$ by quotienting out inversion $\{-1|0\}$. Thus, if we can show that the inversion symmetry is still somehow present in all the bond equivalence classes, then this will imply symmetry enhancement to the eigensymmetry group out to all shells. Note that this also implies that the cumulative symmetry enhancement, that is the symmetry enhancement including all shells simultaneously, is also always enhanced to the eigensymmetry group. We begin by noting that any arbitrary bond $B$ on this crystal can be denoted as:
\begin{equation}
    B = \begin{pmatrix}
        n & m & l\\
        n' & m' & l'
    \end{pmatrix} 
\end{equation}
where $(n,n,l)$ and $(n',m',l')$ are the endpoints of the bond. The action of $\{-1|0\}$ on $B$ can then be defined as:
\begin{equation}
    \{-1|0\} \begin{pmatrix}
        n & m & l\\
        n' & m' & l'
    \end{pmatrix} = \begin{pmatrix}
        -n & -m & -l\\
        -n' & -m' & -l'
    \end{pmatrix}
\end{equation}
One can show however that the image of $B$ under inversion is, up to lattice translations, equivalent to $B$. Indeed we have that
\begin{align}
    \{-1|0\} B - B \equiv &\begin{pmatrix}
        2n & 2m & 2l\\
        2n' & 2m' & 2l'
    \end{pmatrix} \\
    \text{ or }& \begin{pmatrix}
        n+n' & m+m' & l+l'\\
        n'+n & m'+m & l'+l
    \end{pmatrix}
\end{align}
We get two results because the endpoints of a bond are not ordered, so subtracting two bonds can be done in two ways depending on how the endpoints are paired when taking their difference. We see that in the first choice, there is some non-rigid shift between the endpoints of $\{-1|0\}B$ and $B$, whereas in the second choice, the first and second rows of the left-over bond are equal. This shows that inversion does indeed act trivially on the bonds at any shell on the parent crystal, despite the latter not possessing this symmetry. Physically, this corresponds to a symmetry enhancement of any effective $s$-wave tight-binding model on the 1$c$ Wyckoff position of a crystal with space group symmetry $P4_2nm$. This is depicted in Fig. \ \ref{fig:111-1c4n}.

A similar mechanism is likely at play in other cases where the symmetry never breaks down to the parent space group, although we do not show this here. The general procedure would be to check which symmetries present in the eigensymmetry group but not in the parent space group have an action on all bonds on the lattice that can be reproduced by the parent space group. The number of cases where the symmetry never breaks down is shown in Table \ \ref{tab:overview}.

\listoftables

\section{Tables of enhanced symmetries (not shell order resolved)}
\label{app:tab_enh_shell}

\begin{longtable}
{>{\arraybackslash}p{1cm}|>{\arraybackslash}p{1.2cm}|>{\arraybackslash}p{5.4cm}}
\caption{Table of enhanced symmetries for the Triclinic Crystal Class}\\ 
\toprule
Space Group & Wyckoff &   Enhanced Space Group \\
\midrule
\endfirsthead

\toprule
Space Group & Wyckoff &   Enhanced Space Group \\
\midrule
\endhead
\midrule
\multicolumn{3}{r}{{Continued on next column}} \\
\midrule
\endfoot

\bottomrule
\endlastfoot
          1 &    1$a$ &                2(1$a$) \\
          2 &    2$i$ & 2(2$i$) for all shells \\
          2 &    1$h$ & 2(1$h$) for all shells \\
          2 &    1$g$ & 2(1$g$) for all shells \\
          2 &    1$f$ & 2(1$f$) for all shells \\
          2 &    1$e$ & 2(1$e$) for all shells \\
          2 &    1$d$ & 2(1$d$) for all shells \\
          2 &    1$c$ & 2(1$c$) for all shells \\
          2 &    1$b$ & 2(1$b$) for all shells \\
          2 &    1$a$ & 2(1$a$) for all shells \\
\end{longtable}


\nopagebreak
\begin{longtable}{>{\arraybackslash}p{1cm}|>{\arraybackslash}p{1.2cm}|>{\arraybackslash}p{5.4cm}}
\caption{Table of enhanced symmetries for the Monoclinic Crystal Class}\\ 
\toprule
Space Group & Wyckoff &    Enhanced Space Group \\
\midrule
\endfirsthead

\toprule
Space Group & Wyckoff &    Enhanced Space Group \\
\midrule
\endhead
\midrule
\multicolumn{3}{r}{{Continued on next column}} \\
\midrule
\endfoot

\bottomrule
\endlastfoot
          3 &    2$e$ &       10(2$m$), 3(2$e$) \\
          3 &    1$d$ &                10(1$g$) \\
          3 &    1$c$ &                10(1$d$) \\
          3 &    1$b$ &                10(1$c$) \\
          3 &    1$a$ &                10(1$a$) \\
          4 &    2$a$ &       4(2$a$), 11(2$e$) \\
          5 &    4$c$ &       5(4$c$), 12(4$i$) \\
          5 &    2$b$ &                12(2$c$) \\
\end{longtable}

\clearpage

\begin{longtable}[c]
{>{\arraybackslash}p{1cm}|>{\arraybackslash}p{1.2cm}|>{\arraybackslash}p{5.4cm}}
\caption{Table of enhanced symmetries for the Orthorhombic Crystal Class}\\ 
\toprule
Space Group & Wyckoff &                             Enhanced Space Group \\
\midrule
\endfirsthead

\toprule
Space Group & Wyckoff &                             Enhanced Space Group \\
\midrule
\endhead
\midrule
\multicolumn{3}{r}{{Continued on next column}} \\
\midrule
\endfoot

\bottomrule
\endlastfoot
         16 &    4$u$ &                          16(4$u$) for all shells \\
         16 &    2$t$ &                               47(2$t$), 16(2$t$) \\
         16 &    2$r$ &                               16(2$r$), 47(2$s$) \\
         16 &    2$s$ &                               16(2$s$), 47(2$r$) \\
         16 &    2$q$ &                               47(2$q$), 16(2$q$) \\
         16 &    2$p$ &                               47(2$p$), 16(2$p$) \\
         16 &    2$o$ &                               16(2$o$), 47(2$o$) \\
         16 &    2$n$ &                               16(2$n$), 47(2$n$) \\
\end{longtable}

\clearpage

\begin{longtable}[c]
{>{\arraybackslash}p{1cm}|>{\arraybackslash}p{1.2cm}|>{\arraybackslash}p{5.4cm}}
\caption{Table of enhanced symmetries for the Tetragonal Crystal Class}\\ 
\toprule
Space Group & Wyckoff &                                                        Enhanced Space Group \\
\midrule
\endfirsthead

\toprule
Space Group & Wyckoff &                                                        Enhanced Space Group \\
\midrule
\endhead
\midrule
\multicolumn{3}{r}{{Continued on next column}} \\
\midrule
\endfoot

\bottomrule
\endlastfoot
         75 &    4$d$ &                                                          75(4$d$), 83(4$j$) \\
         75 &    2$c$ &              85(2$a$), 83(1$c$), 123(2$f$), 127(2$d$), 123(1$c$), 125(2$c$) \\
         75 &    1$a$ &                                                         83(1$a$), 123(1$a$) \\
         75 &    1$b$ &                                                         123(1$c$), 83(1$c$) \\
         76 &    4$a$ &                                                     76(4$a$) for all shells \\
         77 &    4$d$ &                                                          84(4$j$), 77(4$d$) \\
         77 &    2$b$ &                       131(2$f$), 132(2$c$), 132(2$d$), 123(1$c$), 131(2$b$) \\
         77 &    2$c$ &              139(2$a$), 134(2$b$), 84(2$c$), 136(2$a$), 86(2$b$), 131(2$c$) \\
\end{longtable}

\clearpage

\begin{longtable}
{>{\arraybackslash}p{1cm}|>{\arraybackslash}p{1.2cm}|>{\arraybackslash}p{5.4cm}}
\caption{Table of enhanced symmetries for the Trigonal Crystal Class}\\ 
\toprule
Space Group & Wyckoff &                                  Enhanced Space Group \\
\midrule
\endfirsthead

\toprule
Space Group & Wyckoff &                                  Enhanced Space Group \\
\midrule
\endhead
\midrule
\multicolumn{3}{r}{{Continued on next column}} \\
\midrule
\endfoot

\bottomrule
\endlastfoot
        143 &    3$d$ &                                  174(3$j$), 143(3$d$) \\
        143 &    1$c$ & 162(1$a$), 164(1$a$), 147(1$a$), 191(1$a$), 175(1$a$) \\
        143 &    1$b$ & 162(1$a$), 164(1$a$), 147(1$a$), 191(1$a$), 175(1$a$) \\
        143 &    1$a$ & 162(1$a$), 164(1$a$), 147(1$a$), 191(1$a$), 175(1$a$) \\
        144 &    3$a$ &                              144(3$a$) for all shells \\
        145 &    3$a$ &                              145(3$a$) for all shells \\
        146 &    9$b$ &                              146(9$b$) for all shells \\
        146 &    3$a$ &                                  148(3$a$), 166(3$a$) \\
\end{longtable}

\clearpage

\begin{longtable}
{>{\arraybackslash}p{1cm}|>{\arraybackslash}p{1.2cm}|>{\arraybackslash}p{5.4cm}}
\caption{Table of enhanced symmetries for the Hexagonal Crystal Class}\\ 
\toprule
Space Group & Wyckoff &                                  Enhanced Space Group \\
\midrule
\endfirsthead

\toprule
Space Group & Wyckoff &                                  Enhanced Space Group \\
\midrule
\endhead
\midrule
\multicolumn{3}{r}{{Continued on next column}} \\
\midrule
\endfoot

\bottomrule
\endlastfoot
        168 &    6$d$ &                                  168(6$d$), 175(6$j$) \\
        168 &    3$c$ &            168(3$c$), 191(3$f$), 175(3$f$), 177(3$f$) \\
        168 &    2$b$ &            177(2$c$), 183(2$b$), 175(2$c$), 191(2$c$) \\
        168 &    1$a$ &                                  175(1$a$), 191(1$a$) \\
        169 &    6$a$ &                              169(6$a$) for all shells \\
        170 &    6$a$ &                              170(6$a$) for all shells \\
        171 &    3$b$ &                                  180(3$c$), 171(3$b$) \\
        171 &    6$c$ &                              171(6$c$) for all shells \\
\end{longtable}

\clearpage

\begin{longtable}
{>{\arraybackslash}p{1cm}|>{\arraybackslash}p{1.2cm}|>{\arraybackslash}p{5.4cm}}
\caption{Table of enhanced symmetries for the Cubic Crystal Class}\\ 
\toprule
Space Group & Wyckoff &                                       Enhanced Space Group \\
\midrule
\endfirsthead

\toprule
Space Group & Wyckoff &                                       Enhanced Space Group \\
\midrule
\endhead
\midrule
\multicolumn{3}{r}{{Continued on next column}} \\
\midrule
\endfoot

\bottomrule
\endlastfoot
        195 &   12$j$ &                                  195(12$j$) for all shells \\
        195 &    6$i$ &      200(6$h$), 207(6$f$), 221(6$f$), 195(6$i$), 215(6$g$) \\
        195 &    6$h$ &                                       195(6$h$), 200(6$g$) \\
        195 &    6$f$ &                 195(6$f$), 200(6$e$), 221(6$e$), 207(6$e$) \\
        195 &    6$g$ &                                       195(6$g$), 200(6$f$) \\
        195 &    3$d$ &                 221(3$d$), 207(3$d$), 195(3$d$), 200(3$d$) \\
        195 &    3$c$ &                 200(3$c$), 221(3$c$), 195(3$c$), 207(3$c$) \\
        195 &    4$e$ &                                       215(4$e$), 195(4$e$) \\
\end{longtable}

\clearpage
\section{Bond complex tables (shell order resolved)}
\label{app:bond_complexes}
\nopagebreak
\include{BC_tables}
\end{document}